\documentclass[journal,12pt,onecolumn,draftclsnofoot]{IEEEtran}
\usepackage[T1]{fontenc}
\usepackage{amsmath,bm,amsfonts,amssymb,graphicx,color,multirow,setspace,xfrac,comment,xcolor}
\usepackage{booktabs}
\usepackage{epstopdf}
\usepackage{mathtools}
\usepackage{lipsum}
\usepackage{cite,citesort}
\usepackage{float}
\usepackage{balance}
\usepackage{gensymb}
\begin{document}
\title{27.5 - 29.5 GHz Switched Array Sounder for Dynamic Channel Characterization: Design, Implementation and Measurements}
\author{Harsh Tataria,~\IEEEmembership{Member,~IEEE}, Erik~L.~Bengtsson,~\\Ove Edfors,~\IEEEmembership{Senior Member,~IEEE}, and Fredrik~Tufvesson,~\IEEEmembership{Fellow,~IEEE}
\thanks{H.~Tataria, O.~Edfors, and F.~Tufvesson are with the Department of Electrical and Information Technology, Lund University, Lund, Sweden (e-mail: \{harsh.tataria, ove.edfors,  fredrik.tufvesson\}@eit.lth.se). E.~L.~Bengtsson is with Sony Research Center, Sony Europe, Lund, Sweden (e-mail: erik.bengtsson@sony.com). At the time of conducting this work, E.~L.~Bengtsson was also with the Department of Electrical and Information Technology, Lund University, Lund, Sweden.}
\thanks{This work was partly sponsored by Sony Europe, Ericsson AB, ELLIIT: A Link\"{o}ping-Lund Initiative on IT and Mobile Communication, as well as Stieftelsen f\"{o}r Strategik Foskning (SSF), Sweden. A small part of this paper was presented at the IEEE International Symposium on Antennas and Propagation (APS) 2020, Montreal, Canada \cite{YING1}. Aspects of the sounder design were also discussed in\cite{BENGTSSON1}, a Ph.D. thesis at Lund University, Sweden.}
\thanks{The authors acknowledge valuable discussions with National Instruments Inc., Sony Semiconductor Solutions, and Ericsson AB during the design of the presented channel sounding system. The authors also acknowledge fruitful discussions with Prof. Andreas F. Molisch at the University of Southern California, USA.}}
\maketitle

\vspace{-48pt}
\begin{abstract}
A pre-requisite for the design of wireless systems is the understanding of the propagation channel. While a wealth of propagation knowledge exists for bands below 6 GHz, the same can not be said for bands approaching millimeter-wave frequencies. In this paper, we present the design, implementation and measurement-based verification of a re-configurable 27.5-29.5 GHz channel sounder for measuring dynamic directional channels. Based on the switched array principle, our design is capable of characterizing 128$\times$256 dual-polarized channels with snapshot times of around 600 ms. This is in sharp contrast to measurement times on the order of tens-of-minutes with rotating horn antenna sounders. Our design lends itself to high angular resolution at both link ends with calibrated antenna arrays sampled at 2$^{\circ}$ and 5$^{\circ}$ intervals in the azimuth and elevation domains. This is complemented with a bandwidth of up to 2 GHz, enabling nanosecond-level delay resolution. The short measurement times and stable radio frequency design facilitates real-time processing and averaging of the received wavefronts to gain measurement signal-to-noise ratio and dynamic range. After disclosing the sounder design and implementation, we demonstrate its capabilities by presenting dynamic and static measurements at 28 GHz over a 1 GHz bandwidth in an office corridor environment.
\vspace{-5pt}
\end{abstract}

\begin{IEEEkeywords}
\vspace{-7pt}
Channel sounder design, dynamic channels, mmWave, propagation measurements, switched arrays.
\end{IEEEkeywords}

\section{Introduction}
\label{Introduction}
The explosive growth of mobile data rates and the number of connected devices is motivating the use of spectrum previously unused for cellular communications \cite{RAPPAPORT1,TATARIA1}. As a result, frequency bands close to the millimeter-wave (mmWave) regime from 24.5-29.5 GHz have received considerable attention  \cite{RAPPAPORT1,TATARIA1,SHAFI2}.\footnote{Following the conventional terminology in the related literature, we denote the frequencies 24.5-29.5 GHz as mmWave frequencies, even though, strictly speaking, they are below the ``true" mmWave band which spans 30-300 GHz (see e.g., \cite{SHAFI2} and references therein).} Spectrum in the 24-28 GHz band has been auctioned in the United States by the Federal Communications Commission, triggering rapid research and developments activities \cite{FCC1}. Other countries in North America, Europe, Asia and Oceania are following the same path. To this end, the Third Generation Partnership Project (3GPP) has integrated mmWave frequencies into Release 15 and onward standardization of fifth-generation (5G) systems \cite{3GPP1,ITU1}. For efficient design, performance assessment and deployment planning of mmWave systems, understanding of the propagation channel by means of real-time measurements is \emph{conditio sine qua non}. This enables us to design suitable channel models which capture the physics of the involved propagation processes, in order to characterize their impact on the resulting system performance. Since mmWave systems suffer from high omnidirectional free-space attenuation, large antenna gains are required to effectively penetrate the transmitted signal \cite{SHAFI2}. Given a common environment, mmWave channels behave relatively differently to frequencies below 6 GHz \cite{SHAFI2}. In particular, the \emph{directional characteristics} of mmWave propagation with beamforming antennas need to be better understood over both temporal and spatial domains. 

In order to measure the channel impulse response, a \emph{channel sounder} needs to be designed. The fundamental principle of a channel sounder is to inject a known waveform into the channel, so that suitable signal processing is able to de-convolve the transmitted waveform out of the received signal, in turn acquiring the impulse response. Sounding waveforms are usually designed in accordance with the sounder type. They range from standard pulse trains \cite{DEPARIS1,DEMIR1}, pseudo noise sequences \cite{CASIOLI1,ZWICK1}, chirp signals \cite{HAKEGARD1,PROKES1}, or multitone sequences \cite{CONRAT1,SMULDERS1}. Steady progress is seen in the literature on channel sounder design at mmWave frequencies. Nevertheless, majority of the existing directional sounding setups are for indoor environments and are based on vector network analyzers (VNAs), which use slow chirp and/or frequency scanning, combined with virtual arrays (mechanical movement of a single element or a set of antennas along a track). Such sounders struggle to capture \emph{dynamic environments} and require a cabled connection between the transmitter and receiver limiting their separation distance  \cite{SMULDERS1,GUSTAFSON1,GUSTAFSON2,FU1,BLUMENSTEIN1,MEDBO1}. For outdoor scenarios, the prevalent method for directionally resolved measurements is based on mechanically rotating horn antennas and measuring the channel impulse response for a fixed pair of transmit and receive orientations \cite{RAPPAPORT1,RAPPAPORT2,HUR1,KO1,MACARTNEY1,DU1,KIM1}. There are two major limitations of such an approach: First is the \emph{non-coherent} nature of the measurements, such that the obtained \emph{angular resolution} is limited by the half-power beamwidth (HPBW) of the horn element. As a result, multipath components (MPCs) arriving from different directions may fall into the antenna beamwidth, yet appear as a \emph{single} component. Thus, it has been difficult to predict the precise number of contributing MPCs in a given propagation environment and the fundamental question on mmWave channels being \emph{sparse} (in the sense of having small number of MPCs) remains unanswered. Secondly, mechanical rotation of the horn element typically requires measurement run times on the order of tens-of-minutes or an hour to cycle through all antenna orientations along the azimuth and elevation domains (with a reasonable granularity, for a given snapshot at a particular location). 

To overcome these limitations, several other systems have been reported in the mmWave propagation literature. For instance, the authors in \cite{DU1} present a continuous-wave-based narrowband sounder operating at 28 GHz, feeding 22 dBm of transmit power into a 10 dBi horn antenna. The receiver comprises of a 24 dBi horn antenna mounted on a rotating platform which allows for a full 360$^\circ$ scan across the azimuth plane in 200 ms with 1$^{\circ}$ angular sampling. While this considerably reduces the measurement time, the single element at the transmitter limits the ability to measure truly dynamic, directional channels. Alternatively, the system presented in \cite{MULLER1,DUPLEICH1} is one of the earliest to consider the possibility of measuring dynamic mmWave multiple-input multiple-output (MIMO) channels. The designed system can facilitate measurements across multiple frequency bands with a variable bandwidth of 3.5 GHz or more, and utilizes dual-polarized horn antennas at both transmit and receive link ends. Examples of the power azimuth and elevation spectra at 60 GHz are presented in \cite{MULLER1}. Nevertheless, having dual-polarized horn antennas hugely constrains the total field-of-view of the transmitter and receiver. The authors in \cite{PAPAZIAN1,SUN1} have proposed a sounder based on an \emph{array} of directional horn antennas combined with fast switching at the receiver with a single transmit antenna. On the other hand, studies in \cite{SALOUS1,SALOUS2} present a multi-band MIMO channel sounder operating at 30 GHz, 60 GHz and 90 GHz with 8$\times$8 and 2$\times$2 antenna arrays, respectively. The MIMO operation is realized by employing switches at the intermediate frequency (IF) along with parallel frequency conversions. Nonetheless, the output transmit powers are limited to 16 dBm at 30 GHz, 7 dBm at 60 GHz and 4 dBm at 90 GHz, greatly reducing the dynamic range of the sounder. The authors in \cite{WEN1} present a sounder with 4 transmit antennas multiplexed with a single pole four throw (SP4T) switch and 4 receive antennas with 4 parallel down-conversion chains. Similar to \cite{SALOUS1}, the transmit power is limited to 24 dBm. Collectively, since majority of the aforementioned works use the switched array principle, the MIMO order is limited by a limited angular resolution, not being able to fully compliment the delay resolution on offer. Furthermore, for dynamic environments, the vast majority of the reported measurements \emph{either} investigate the angular characteristics without considering the temporal behavior of parameters, or focus on the temporal behavior without considering the angular characteristics - see e.g., \cite{HE1,MACCARTNEY2}. The authors of \cite{CAI1} describe a VNA-based system for tracking the MPC parameters in indoor scenarios. To the best of our knowledge, the sounder presented in \cite{BAS1} is the most complete thus far in terms of characterizing the amplitude, delay, angular and Doppler properties of the mmWave channel in dynamic outdoor scenarios. The system is based on the concept of a phased array and performs fast beam switching/steering (on the order of 2 $\micro$s) in both the azimuth and elevation domains with 8$\times$2 arrays at the transmitter and receiver. However, this setup is restricted to measure over 90$^\circ$ sectors in azimuth and elevation, limiting the overall sounding field-of-view and hence MPC determination. Furthermore, in order to maintain the phase coherency across multiple radio-frequency (RF) down/up-conversion chains, the operational bandwidth was constrained to 400 MHz, which  directly limits the delay resolution. The angular resolution is also restricted by the size of the arrays used at both ends with the active radiation patterns in \cite{BAS1}.   

Unlike all previous systems, we present a re-configurable 27.5-29.5 GHz channel sounder capable of measuring 128$\times$256 dual-polarized channels from the transmitter to the receiver. Our design facilitates measurements of an order-of-magnitude more channels with an equivalently higher angular resolution in comparison with other sounders presented in \cite{RAPPAPORT1,RAPPAPORT2,HUR1,KO1,MACARTNEY1,DU1,KIM1,MULLER1,DUPLEICH1,PAPAZIAN1,SUN1,SALOUS1,SALOUS2,WEN1,HE1,MACCARTNEY2,BAS1,SUN2,GENTILE1,ALALAURINAHO1,ZHANG1}. Our system is catered to extract dynamic directional behavior of channels with an overall field-of-view of 180$^\circ$ at the transmitter and 360$^\circ$ at the receiver. Our design is based on the classical switched array principle, where a network of high power, high speed SP4T switches are cascaded and interfaced with both transmit and receive antenna elements (exact structure discussed later on). This is followed by a single 27.5-29.5 GHz up/down-conversion RF chain to/from baseband. With the transmit and receive switch control implemented in dedicated field programmable gate arrays (FPGAs) at the two ends, our design is capable of switching from one element to the next in approximately 18 $\micro$s inclusive of the switching time \emph{and} processing latency. This relies on the fact that 4 complex waveforms are averaged at the receiver and therefore is re-configurable. To this end, a complete MIMO snapshot of 32768 (128$\times$256) channels can be measured in around 600 ms, achieving the same overall effect as the rotating horn antenna sounders with several orders-of-magnitude reduction of measurement time for a given measurement location. While our system supports a maximum bandwidth of 2 GHz, via software tuning, we are also able to re-configure the bandwidth to 1 GHz or below. 

Relative to most other systems, our setup has six distinguishing features; (1) With high angular \emph{and} temporal resolutions, we are able to accurately characterize MPC amplitudes, delays, angles, Doppler and polarization parameters of the dynamic mmWave channel. (2) To get additional accuracy in parameter estimation, non-sequential switching of the 32768 channels is implemented via a fixed codebook known to the transmitter and receiver (details provided later in the paper). (3)  Via careful RF design, negligible phase drift is seen between the reference local oscillators (LOs) at both link ends, even in the absence of a cabled connection for timing synchronization. This enables averaging of multiple complex waveforms to enhance the measurement signal-to-noise ratio (SNR) - something which is particularly valuable for measurements which require a large dynamic range with high geometric attenuation. (4) The active antenna radiation patterns at the transmit and receive arrays with the cascaded switching network and interconnects are \emph{calibrated} over a 4 GHz band from 26-30 GHz, with angular sampling of 2$^\circ$ in the azimuth and 5$^\circ$ in the elevation. This granularity in angular sampling is superior by an order-of-magnitude relative to most other systems, and is otherwise considered to be extremely difficult and time consuming to achieve at mmWave frequencies. (5) The calibrated amplitudes and phases of each element enables us to utilize high-resolution parameter extraction algorithms, such as RIMAX \cite{RICHTER1,TATARIA2} to maximize the estimation accuracy of propagation parameters. (6) Our design demonstrates the highest degree of \emph{re-configurability}, allowing us to manipulate, in real-time, the number of measured channels, averages per-channel, polarization modes, bandwidth and center frequency. 

Before presenting the contributions of the paper, we make one further remark regarding the various channel sounding architectures seen in the literature. Rather interestingly, architectures based on switched arrays can be further classified into \emph{switched beam architectures} (SBA), \emph{switched horn antenna arrays} (SHA), and \emph{switched patch arrays} (SPA), respectively  \cite{RAPPAPORT1,RAPPAPORT2,HUR1,KO1,MACARTNEY1,DU1,KIM1,MULLER1,DUPLEICH1,PAPAZIAN1,SUN1,SALOUS1,SALOUS2,WEN1,HE1,MACCARTNEY2,BAS1,SUN2,GENTILE1,ALALAURINAHO1,ZHANG1}. The major difference between these approaches relates to the achievable dynamic range and scaliability of the design. The SBA achieves high gain by coherently combining signals from multiple elements, while the SHA achieves large array gains leveraging the horn antenna directivity. In contrast, the SPA digitizes the channel impulse response measured at each antenna to achieve array gain in the digital domain via further processing. For the SBA, a critical part of the circuit design is the phase shifting network. Using standard complementary metal oxide semiconductor or gallium arsenide technology, an insertion loss on the order of 10 dB is expected with a moderate switching rate of 1 $\micro$s \cite{SHIN1}. This is in the order of a waveform duration, yet low enough to not significantly degrade the dynamic measurement capability. Thus, to fully exploit the array gain and overcome the insertion loss of the network, an amplifier (low-noise at the receiver or high power at the transmitter) between the element and the phase shifter is needed. This has a significant impact on the power consumption, implementation complexity and cost when the number of elements are increased. Moreover, they also set a fundamental practical limit to the beam switching rate, which should be smaller than the overall channel acquisition time. The main limitation of SHA relates to cost, mechanical complexity and size, since each well designed horn element is expensive and needs to be arranged mechanically to cover a unique angle (see e.g., the array known as ``Porcupine" used in \cite{ATNT1}, which is about 0.5 m in diameter). Different to SBA and SHA, the SPA design has the ability to utilize the recent advances in RF switching technology, significantly reducing the net insertion loss to less than 2 dB for a single switch with a switching rate on the order of 100 ns at mmWave frequencies \cite{PS1}. This has opened up the prospects for massive SPA designs, such as the one presented in this paper, offering the best trade-off solution in terms of implementation complexity, cost and performance. The contributions of the paper can be summarized as follows: 
\begin{itemize}
    \item We present the design, implementation and verification of a re-configurable switched channel sounder operating over 27.5-29.5 GHz, capable of measuring 128$\times$256 dynamic directional dual-polarized channels in around 600 ms. Relative to state-of-the-art systems, the design provides superior angular resolution and is complimented with a nanosecond delay resolution. As such, the design presents an opportunity to \emph{jointly} estimate and track the MPC parameters. Also in contrast to other systems, a 360$^\circ$ field-of-view at the receiver is maintained, while sounding the channels from a 180$^\circ$ sector at the transmitter, increasing the likelihood of measuring MPCs from a wide angular range. 
    \item We measure and characterize the active radiation patterns of each transmit and receive element in an anachoic chamber with dense angular sampling of 2$^{\circ}$ in the azimuth and 5$^\circ$ in the elevation, across a 4 GHz frequency band from 26-30 GHz having 250 MHz frequency steps. The antenna measurements were carried out with the integrated switching and interconnect network. This is a complex task, where the resulting characterization time scales with the number of elements, density of angular sampling, measurement bandwidth and frequency steps. As such, the 256 element array resulted in measurement time of 55 hours, while the 128 element array resulted in 29 hours to measure.  
    \item We present sample measurements of dynamic and static propagation channels at a center frequency of 28 GHz over a 1 GHz bandwidth in an indoor office corridor environment. We extract the directional propagation parameters across multiple time instances and spatial locations via the RIMAX high-resolution algorithm \cite{RICHTER1}. Our results show the appearance and disappearance of MPCs as a function of the scatterer mobility, and we study the directional delay, angular and power characteristics in both dynamic and static scenarios. The short MIMO snapshot times are complemented by highly stable RF design enabling efficient FPGA implementation to carry out complex real-time averaging of multiple received waveforms, increasing the measurement SNR. Unlike previous systems, approximately 1 Gb of measured data is efficiently processed and downloaded per-second. 
\end{itemize}

\section{Channel Sounder Design and Implementation}
\label{ChannelSounderDesign}
\vspace{3pt}
The general principle is as follows: A known waveform is generated at baseband and up-converted to carrier frequency of interest, followed by amplification and transmission. On the receive side, the waveform is down-converted, sampled and stored for post-processing. The channel impulse response is extracted from the received signal envelope relative to what was transmitted. In order to extract the directional characteristics of the channel, \emph{multiple repetitions} of the sounding waveform are transmitted into (and hence received from) different directions, illuminating different transmit and/or receive antennas, followed by high resolution processing.  
\vspace{-13pt}
\subsection{Sounding Waveform Design}
\label{SoundingWaveformDesign} 
In line with the setups in \cite{CONRAT1,PAPAZIAN1,BAS1}, the baseband sounding waveform implemented at the transmitter's host interface is a multi-tone \emph{Zadoff-Chu (ZC)} sequence. ZC sequences have ideal correlation properties in both time and frequency domains, and are thus well suited for channel sounding. In contrast to other waveforms, they are also easily scalable across both time and frequency domains. Naturally, wider \emph{signal bandwidths} facilitate higher frequency selectivity, while the \emph{total duration} of the sounding waveform is designed keeping in mind the furthest desired resolvable MPC.\footnote{Naturally, in the time domain, this gives an estimate of the maximum excess delay of the measured channel impulse response.} Denoted by $x(t)$, the sounding waveform can be written as $x(t)=\sum\nolimits_{n=-N}^{N}
\hspace{3pt}e^{\hspace{1pt}j\left(2\pi{}n\delta{}\hspace{-1pt}ft+\theta_{n}\right)}$. The total number of tones are given by $2{}\hspace{-1pt}N+1$, while $\delta{}\hspace{-2pt}f$ denotes the tone spacing, and $\theta_n$ is the complex phase of the $n$-th tone. In our design, a total of 2002 tones each spaced by 500 kHz were selected to cover the instantaneous measurement bandwidth of 1 GHz.\footnote{We note that for the general case of measurement bandwidths up to 2 GHz, we proportionally increase the total number of tones with the inter-tone spacing fixed to 500 kHz. This is a software re-configurable parameter in the designed system.} Following the methodology presented in \cite{BAS1,FRIESE1}, the values of $\theta_n$ are designed to optimize the peak-to-average-power ratio (PAPR) of $x(t)$ for power efficient signalling to maximize the forward link SNR. Considering the above, we implement a single ZC sequence in the form of $x(t)$, which composes of 2048 samples spanning 2.6 $\mu$s in time, sampled with a rate of 1.4 GSamples/s. The resulting frequency spectrum of the ZC signal is flat across the desired bandwidth of 1 GHz (re-configurable up to 2 GHz when applicable), providing the required SNR uniformity across all frequency tones.
\begin{figure}[!t]
\vspace{-5pt}
\centering
\includegraphics[width=9.4cm]{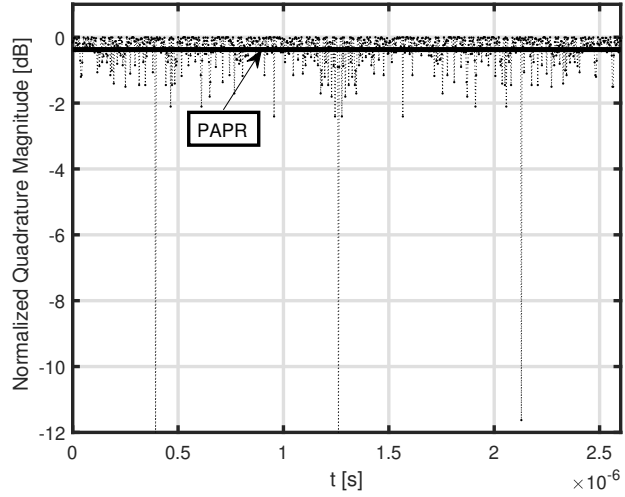}   
\caption{Normalized quadrature magnitude of the ZC sequence in dB as a function of time with a PAPR of 0.349 dB.}
\label{ZCSeqQuadMagnitude}
\vspace{-18pt}
\end{figure}
Figure~\ref{ZCSeqQuadMagnitude} depicts the normalized quadrature magnitude of $x(t)$, which is seen to have a PAPR within 0.35 dB, enabling us to transmit with power as close as possible to the 1 dB compression point of the transmit power amplifier (PA), operating in the linear region of its characteristics curve. For greater robustness against measurement noise, \emph{multiple} ZC waveforms are transmitted in real-time (more information on this is presented later in the text). At the receiver, the multiple received replicas are averaged and correlated with the transmitted ZC waveform. We present further details of this in the subsection that follows. 
\vspace{-12pt}
\subsection{Sounding Frame Structure}
\label{SoundingFrameStructure} 
\begin{figure}[!t]
\centering
\vspace{-10pt}
\includegraphics[width=11cm]{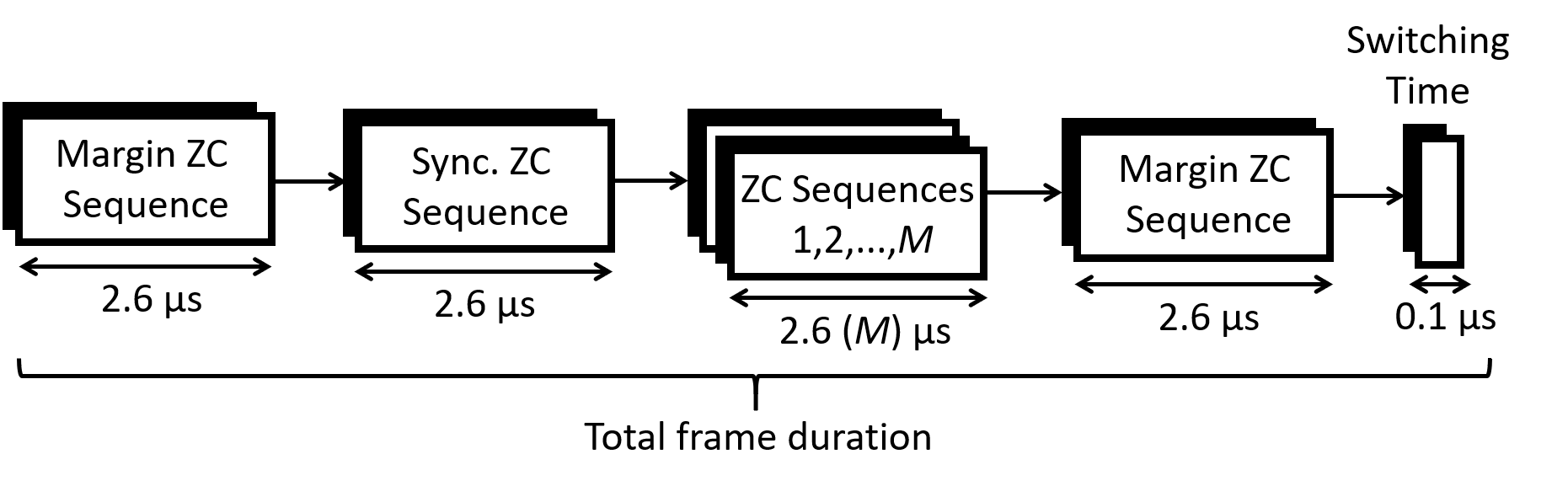}  
\vspace{-14pt}
\caption{Frame structure composing of two margin sequences, $M$ core sequences, one sync. sequence and switch settling time.}
\label{FrameStructure}
\vspace{-24pt}
\end{figure}

\vspace{-3pt}
As will be described in greater details later in the paper, the sounder is designed with a 128 element uniform planar array (UPA) at the transmitter and a 256 element cylindrical array at the receiver supporting dual polarization. Both arrays are equipped with integrated RF switches to switch through each antenna combination from transmitter to the receiver. According to the switched array principle, both link ends are equipped with a single RF up/down-conversion chain. From each transmit antenna, the \emph{frame structure} is built up from multiple repeated ZC sequences, followed by a \emph{guard period} for settling the RF switches after activation. The guard duration of the RF switches is aligned with the 99\% settling time of 100 ns for most commercial RF switches operating up to 44 GHz \cite{PS1,AD1}. The number of ZC sequences transmitted from each antenna is re-configurable, and is a design parameter to strike the right balance between the total time taken to measure one antenna combination and robustness against measurement noise. In addition, the first and the last ZC sequence in each frame are added as \emph{margins} in the case there are random drifts from the local oscillator (LO) at the receiver. The second ZC sequence of the frame is used by the receiver as a margin in case if arbitrary jitter from the clock source (CLK) take place which could cause the receiver to be out of synchronization from the transmitter (especially for calibrated CLKs used independently at both link ends). The remaining arbitrary number of ZC sequences are used for real-time averaging.  Since each ZC sequence is of duration 2.6 $\mu$s, the frame duration with \emph{four} ZC sequences for real-time averaging is given by 18.3 $\mu$s. To this end, with 128 transmit and 256 receive antennas, 32768 channel combinations yield a MIMO snapshot that can be measured in around 600 ms. Naturally, a lower number of core ZC sequences used for real-time acquisition reduces the duration of the MIMO snapshot. In the general case, the frame structure for any transmit-receive antenna combination with the number of core ZC acquisitions set to $M$ is depicted in Fig.~\ref{FrameStructure}.  

As our aim is to accurately extract the propagation parameters in dynamic scenarios, we need to ensure that the duration of one MIMO snapshot is \emph{shorter} than the channel coherence time. This implies that the MIMO cycle rate - the inverse of the duration between two adjacent snapshots should be greater than or equal to \emph{half} of the maximum absolute Doppler shift, in order to avoid ambiguities in Doppler frequency estimation of MPCs. In a switched sounder, since the MIMO snapshot duration increases with the number of antenna combinations, there is an inherent conflict between the desired accuracy of the angle-of-arrival (AOA) and angle-of-departure (AOD) estimates (which implicitly demand higher numbers of antenna combinations) and the maximum admissible Doppler frequency. The authors of \cite{YIN1} were first to report that the choice of \emph{sequential} antenna switching sequence was causing this limitation. Instead, a \emph{randomized} (non-sequential) switching sequence has the potential to greatly extend the Doppler estimation range and nullify the aforementioned ambiguities. This principle relies on the fact that the signal phase can be inversely derived from the Doppler spread, as long as its second-order statistics are time invariant, and that aliasing from systematic spatial switching can be avoided since each direction is given equal importance. In a similar line, with directional antennas, authors of \cite{PEDERSEN1,PEDERSEN2} introduced the use of the normalized sidelobe levels as the objective function to derived the necessary conditions of array switching sequence which leads to ambiguities. Hence, similar to \cite{YIN1,WANGR1}, we employ the use of \emph{pseudo-random} switching at both link ends of our sounder. From an implementation viewpoint, both sides of the sounder locally generate switching sequences via software in a codebook that is known to both ends prior measuring.  

The same frame structure holds for measurements of dynamic polarization properties, which is one of the unique features of the sounder relative to previous setups. Since the polarization properties of MPCs is defined by the \emph{phase offset} between vertical and horizontal components, ideally both polarizations need to be measured simultaneously. For measurements in dynamic environments, both the time offset between the measurements and the estimated Doppler spread need to be considered as design parameters in the estimation of the polarization properties. By switching the antenna elements, and hence polarizations, randomly while keeping the opposite polarized ZC sequences together, we can minimize the phase offset error of not estimating both polarizations simultaneously down to the order of one degree.\footnote{This has been checked via numerical simulations of the polarization ambiguity function with the exact switching sequences employed in our sounder, but not presented in the paper due to space reasons. Similar methodologies can be found in \cite{WANGR1}.}   
\vspace{-14pt}
\subsection{Transmit and Receive Architectures}
\label{TransmitandReceiveArchitectures}
\vspace{-4pt}
We now present details on the designed antenna arrays and switching architecture, antenna array calibration, transmit and receive hardware, and real-time host as well as FPGA software implementation aspects. In order to maximize clarity, we distribute the aforementioned items into the different subsections below.  
\subsubsection{Antenna Arrays and Switching Architecture}
\label{AntennaArraysandSwitchingArchitecture}
\begin{figure}[!t]
\begin{center}
\includegraphics[width=15.5cm]{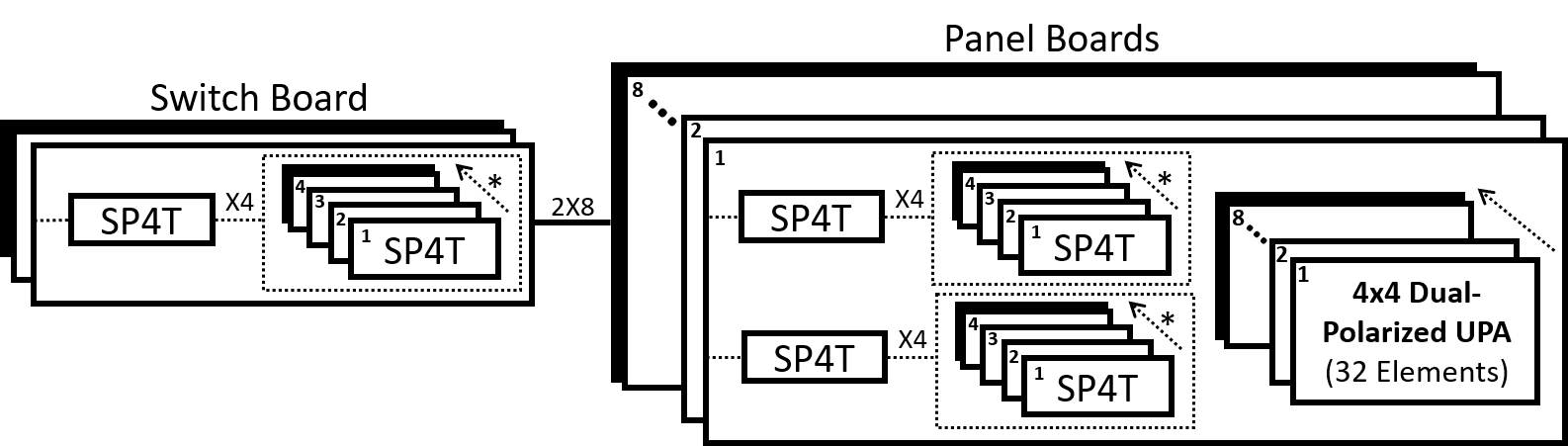}
\vspace{3pt}
\caption{The cascaded switching topology divided into a switch board feeding up to 8 antenna panel boards each with 16 dual-polarized antenna elements. This enables control of 256 elements. The same architecture is used for the 128 element array with 4 antenna panels instead of 8. Note that * in the figure denotes stacked SP4T switches.}
\label{SwitchingArchitecture}
\vspace{-7pt}
\end{center}
\vspace{-22pt}
\end{figure}

\begin{figure}[!t]
\begin{center}
\vspace{-10pt}
\includegraphics[width=13cm]{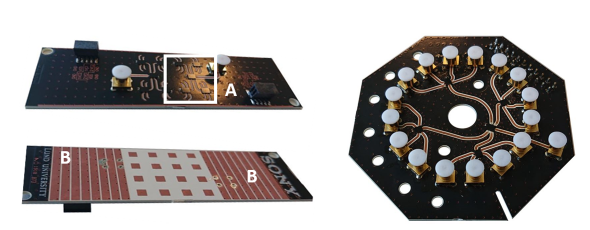}  
\vspace{-5pt}
\caption{Left-hand side: Fabricated 4$\times$4 dual-polarized UPA panel with integrated switching. The rear part of the array containing the switching topology is demonstrated in the top half of the figure, while the bottom half contains the front of the array panel with the three-layered element design and corrugation structure for surface current minimization. Right-hand side: Switchboard with two layers of cascaded SP4T switches, coaxial connectors for input and 16 outputs with a control interface.}
\label{SinglePanelandSwitchBoard}
\end{center}
\vspace{-30pt}
\end{figure}

The antenna arrays employed at both the transmit and receive ends are composed of dual-polarized patch elements, and are interfaced with a quadruple cascaded RF switching network. The switching circuitry needed to facilitate the design of the transmit/receive arrays is based on custom made 28 GHz SP4T switches donated by Sony Semiconductor Solutions. On average, the switches can sink 32 dBm and have a net insertion loss on the order of 1.5 dB - well exceeding the design specifications of commonly available commercial RF switches in the public domain. In general, the switching architecture enables control of up to 256 elements according to the pre-defined pseudo-random switching sequence, and its design is divided in two parts: The first half is located on a dedicated \emph{switch board} that carries two levels of cascaded switches and distributes the input signal to 16 coaxial connectors. The switch board is interfaced with the second half of the switching network, which is located at the backplane of individual \emph{antenna element panels} (referred to as panel boards later in the text), where each panel has two coaxial inputs that are switched out to 32 antenna feeds compiled in a 4$\times$4 UPA supporting two polarizations. For the 256 element receive array, there are eight such panels in an \emph{octagon}, while for the 128 element transmit UPA, there are four panels in a \emph{rectangle}. The overall cascaded switching topology supporting a maximum of 256 elements is demonstrated in Fig.~\ref{SwitchingArchitecture}. 
Both the switch board and antenna panel boards are designed with six layer printed circuit board (PCB) technology based on Rogers RO4450B substrate with optimized loss tangent ($\tan(\tilde{\delta})$ = 0.004), dielectric constant ($\epsilon_r$ = 3.54$\pm$0.05), and above average thermal conductivity (0.6 - 0.8). The six layers of the PCB spans 1020 $\mu$m of width in total. The process supports stacked vias through all six layers.\footnote{We note that stacked vias enable signal paths to go through arbitrary layers of a PCB without introducing sidewise offsets.} This enables a more compact layout with low insertion losses and has facilitated half of the switching circuitry to be at the back of the panels next to the feed of the patch elements. The front and backplanes of a single 32 element UPA panel are depicted in the left-hand side of Fig.~\ref{SinglePanelandSwitchBoard}. Note that the corrugation structure (horizontal stripes marked with the letter ``B") is visible on the front view of the antenna panel above and below the patch elements. This was introduced to prevent surface currents from distorting the antenna radiation patterns and to minimize general energy spillover. However, due to manufacturing processing limitations, the required grounding associated with the corrugation structure could not be supported from a mechanical strength perspective. Instead, a layer of absorbing material (carbon induced rubber) was added, which will be visible later. 

Each patch antenna seen on the left-hand side of Fig.~\ref{SinglePanelandSwitchBoard} is designed as a \emph{coupled parasitic resonator}, where the feed to the element is connected through stacked vias to the switch outputs (shown with a square with marker ``A") at the panel backplane. The elements are designed in a three-layered structure: The lowest layer is the ground plane; the second layer elevated to 100 $\mu$m is a dual-feed coupling element; and the top layer at 300 $\mu$m, is the radiator. The coupling element in combination with the radiator generates two closely spaced resonances that are tuned to achieve a bandwidth that covers the desired measurement frequencies. 
\begin{figure*}[!t]
\centering
\includegraphics[width=10.8cm]{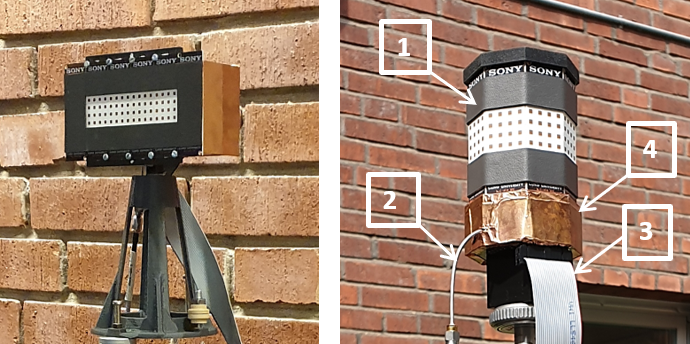}
\vspace{3pt}
\caption{Left-hand side: 128 element UPA for the transmitter; Right-hand side: 256 element octagonal array for the receiver. Both arrays contain the discussed integrated switches and control distribution networks.}
\label{Arrays}
\vspace{-20pt}
\end{figure*}
We note that each element has been designed to deliver a HPBW in the azimuth of 85$^{\circ}$ and elevation of 50$^{\circ}$. These are retained across both vertical and horizontal polarizations. In Sec.~\ref{ArrayCalibration}, we present details of antenna radiation pattern  characterization and confirm the aforementioned HPBW values via measurements. For stable RF design, each element's associated port is isolated by 20 dB with a -10 dB return loss over 26-30 GHz. The total insertion loss from the RF interconnects, cabling and switches is approximately 15 dB for an antenna. Each switchboard interface includes the first two layers of the switch cascade, and is equipped with a coaxial connector for feeding 16 coaxial connectors for signal distribution to the respective panels. This is demonstrated on the right-hand side of Fig.~\ref{SinglePanelandSwitchBoard}. Using both the panel and switch boards, the transmit and receive arrays are shown in Fig.~\ref{Arrays}. As mentioned earlier, the corrugation structure on the panel boards of both arrays is overlaid with absorbing material (black) to minimize surface currents and leakage. On the receive array, this is marked by ``1" on Fig.~\ref{Arrays}. The RF cable to/from the receive array is marked with ``2". The 22-pol 8-bit transistor-transistor logic (TTL) control (cable) for antenna switching (maximum 256 states) is depicted by ``3" on the, while the copper shielding for isolating the active control components suppressing unwanted radiation is shown with ``4" on the receive array. The same trends can be observed for the transmit array with the exception of the shielding behind the array due to the location of the control blocks. Both arrays are further filled with an absorber to avoid the creation of a cavity with resonances potentially leaking out.
\subsubsection{Antenna Array Calibration}
\label{ArrayCalibration}
The three-dimensional gain and phase patterns over both polarizations for both antenna arrays was characterized in an anechoic chamber of dimensions 9.2 m $\times$ 5.3 m $\times$ 5.2 m (length $\times$ width $\times$ height). The arrays were calibrated across 26-30 GHz (4 GHz bandwidth) with a frequency step size of 250 MHz. In order to maximize the sensitivity of the arrays when used with high-resolution post-processing, characterization of both arrays was done with dense angular sampling of 2$^\circ$ in the azimuth and 5$^{\circ}$ in elevation. The measurements utilize the principle of a \emph{compact antenna test range (CATR)}, where a source and feeder system transmit spherical wavefronts with power 10 dBm at a given frequency within the 4 GHz bandwidth at a particular time instance. A passive reflector is used to collimate the spherical wave into uniformly homogeneous plane waves for illuminating the fabricated arrays, a.k.a. device under test (DUT) over a distance substantially shorter than in a comparable far-field range. Since the measurements utilized the classical \emph{roll-over azimuth} principle, the DUTs were mounted on a rotating arm which was controlled by a programmable positioner. The element switching (including two polarization states) for both DUTs was controlled by a host computer (PC) interfacing with a switch control circuit leading towards the 8-bit TTL signals utilized by RF switches on the backplane of the antenna panels. Note that the distance from the source to the reflector was 4.58 m, while the reflector was placed 5.50 m away from the DUTs. The mounting height of the DUTs and the height of reflector's geometrical center from the base of the chamber was 3.40 m. To separate the antenna characteristics from the rest of the transmitter/receiver circuitry, the DUTs were calibrated without the PA or low-noise amplifier (LNA). Together with the large number of elements, the dense sampling of the field in azimuth and elevation with the wide measurement bandwidth yielded an unusually high number of physical movements of DUTs. This contributed to a total time of 55 hours to characterize the 256 element array, while 128 element array resulted in 29 hours of measurement time. 

\begin{figure}[!t]
\vspace{-15pt}
    \centering
    \hspace{-20pt}
    \includegraphics[width=15cm]{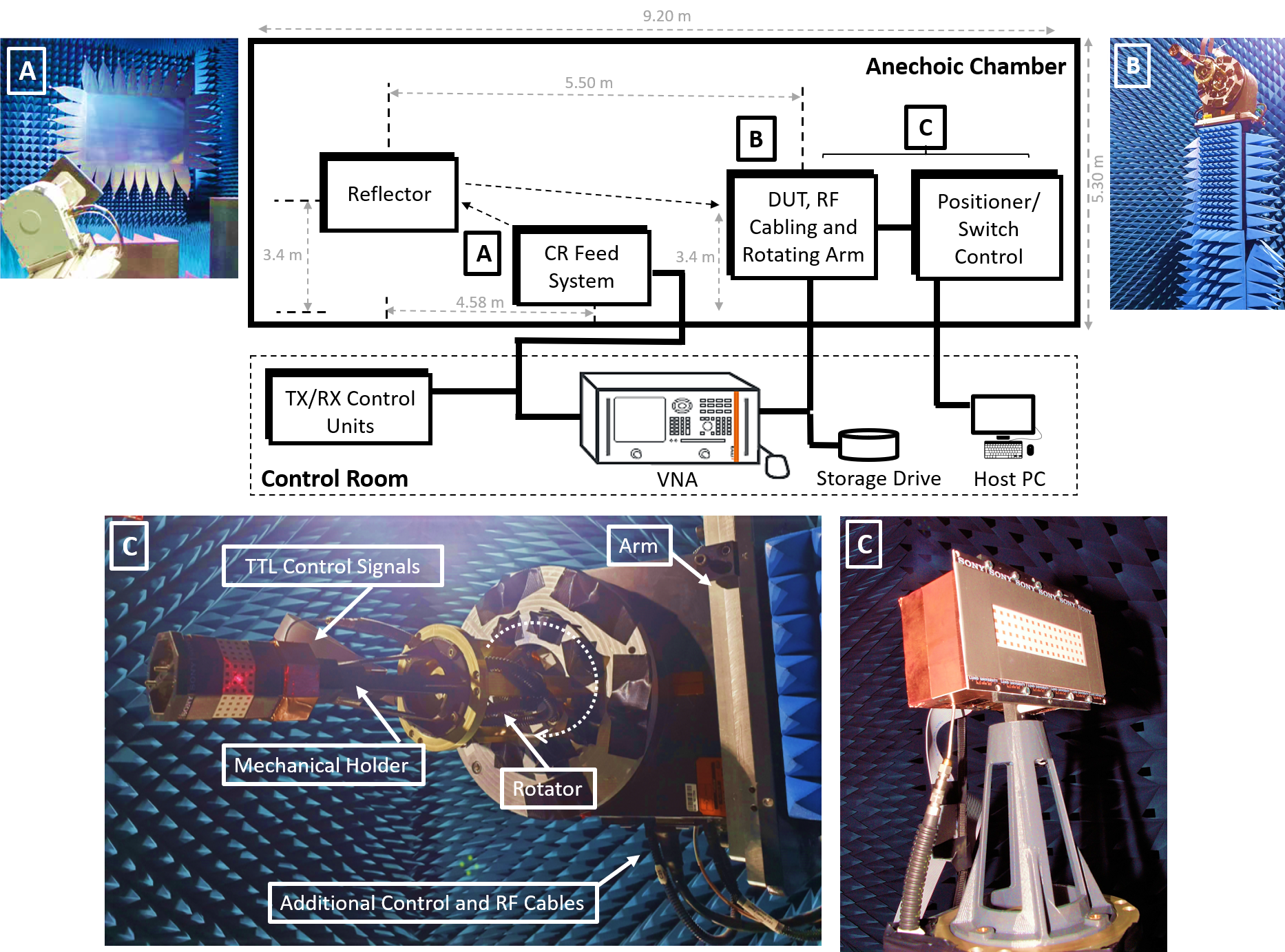}
    \vspace{-3pt}
    \caption{Figure center-top: Anechoic chamber measurement setup for characterization of the per-element gain and phase patterns of the fabricated transmit and receive antenna arrays from 26-30 GHz. The letters ``A", ``B" and ``C" each map to the sub-figures depicted towards the left-hand side, right-hand side and below the central figure. The term ``TX/RX Control Units" refers to transmit/receive control as well as some signal conditioning functionality required by the measurement software system.}
    \label{AntennaMeasurementSetup}
    \vspace{-6pt}
\end{figure}

\begin{figure}[!t]
\vspace{-10pt}
\hspace{10pt}
    \includegraphics[width=15cm]{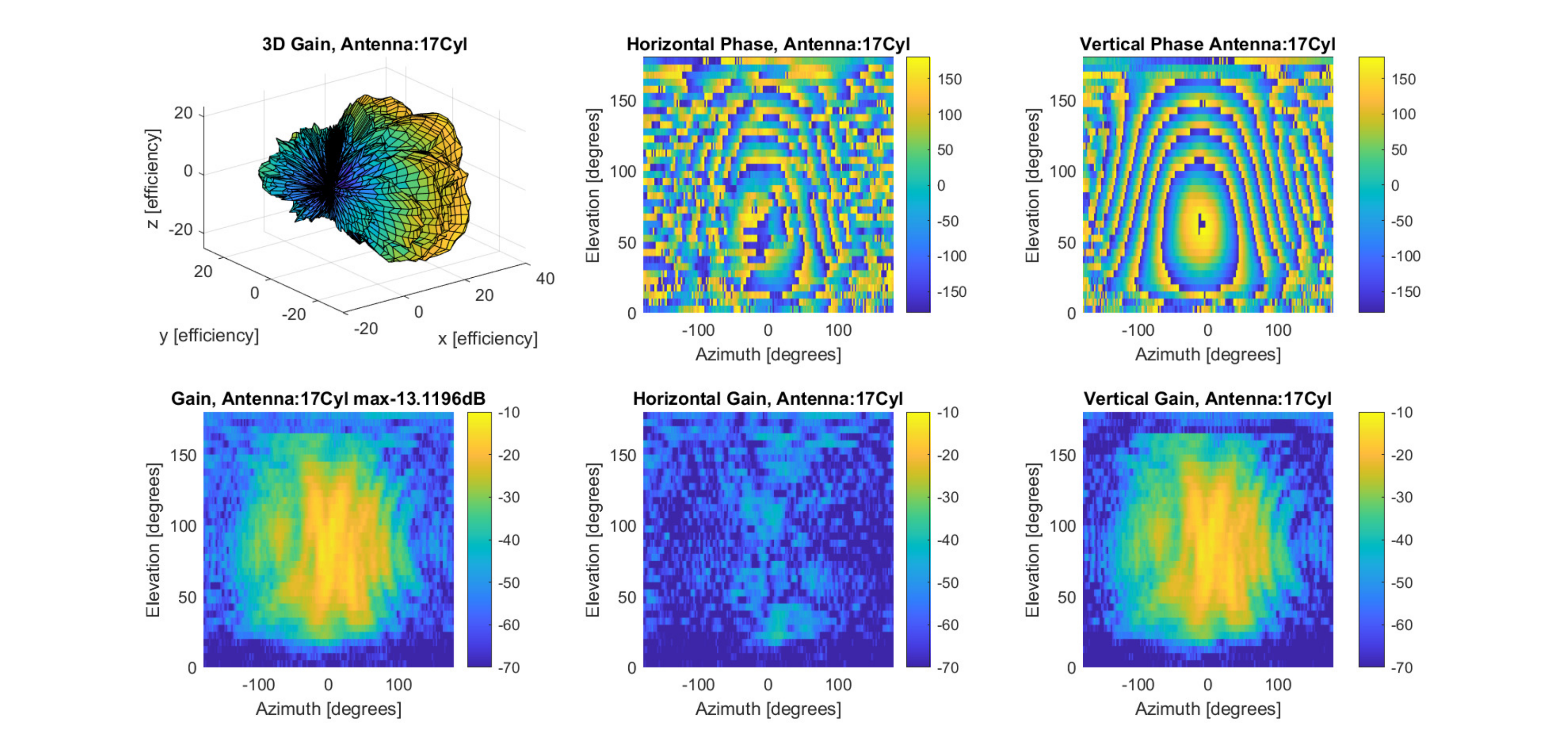}
    \vspace{-15pt}
    \caption{Measured three-dimensional gain and phase radiation pattern of element 17 of the receive array at 28 GHz. The color bars are represented in dB and degrees where applicable.}
    \label{RXAntenna17Pattern}
    \vspace{-25pt}
\end{figure}

\begin{figure}[!t]
    \hspace{10pt}
    \includegraphics[width=15cm]{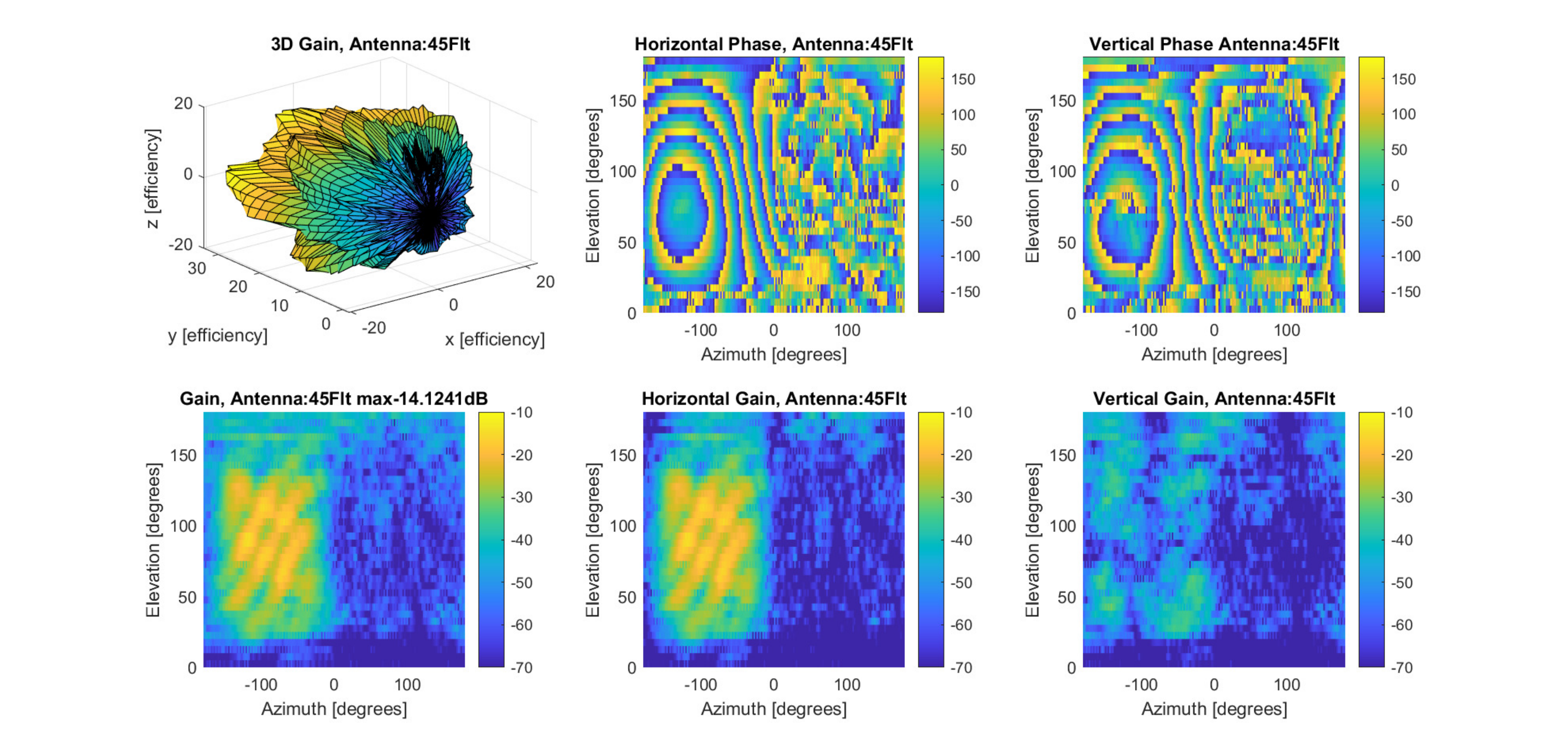}
    \vspace{-17pt}
    \caption{Measured three-dimensional gain and phase radiation pattern of element 45 of the transmit array at 28 GHz. The color bars are represented in dB and degrees where applicable.}
    \label{TXAntenna45Pattern}
    \vspace{-24pt}
\end{figure}

Figure~\ref{AntennaMeasurementSetup}, the center-top part of the figure, demonstrates the overall measurement setup (inside and outside the chamber) for calibrating the DUTs. The letters ``A", ``B", and ``C" each map to sub-figures shown towards the left, right and below the central sub-figure. Here the CRAT source feed system, reflector and both DUTs are depicted.  
After performing the measurements, the element gain and phase patters in both polarizations across the measurement bandwidth were analyzed. Figures~\ref{RXAntenna17Pattern} and \ref{TXAntenna45Pattern} show example patterns at 28 GHz from antenna element 17 at the receive array and from antenna element 45 at the transmit array, respectively. For both figures, one can extract several trends: Firstly, the radiation efficiency is maximum towards the broadside direction of the elements seen from examining the top and bottom-left-hand sub-figure of both Figs.~\ref{RXAntenna17Pattern} and \ref{TXAntenna45Pattern}. Secondly, for the illuminated polarization mode, the earlier quoted azimuth and elevation HPBWs of 85$^{\circ}$ and 50$^{\circ}$ are confirmed. These can be seen by closely examining either the ``horizontal" or the ``vertical gain" of the individual patterns presented in both figures. Thirdly, the cross-polarization discrimination ratios (i.e., the power present in the illuminated polarization relative to the non-illuminated polarization) is approximately 30 dB, and is retained across both array elements.   
For the non-illuminated polarization mode, the element phase retains its overall shape but is expectantly distorted relative to the phase of the illuminated polarization. This can be observed by examining the sub-figures which contain sub-titles of ``Horizontal Phase" and ``Vertical Phase", respectively. Although not shown here due to space constraints, the remaining elements measured gain and phase patterns for both arrays show equivalent trends as those described above. Uniform gain and phase performance is observed over the 4 GHz measurement bandwidth. We now describe the transmit and receive hardware operation of the designed sounder. 
\subsubsection{Sounder Operation With Transmit and Receive Hardware}
\label{Sounder Operation With TransmitandReceiveHardware}
\begin{figure}[!t]
    \vspace{-15pt}
    \centering
    \hspace{10pt}
    \includegraphics[width=12.4cm]{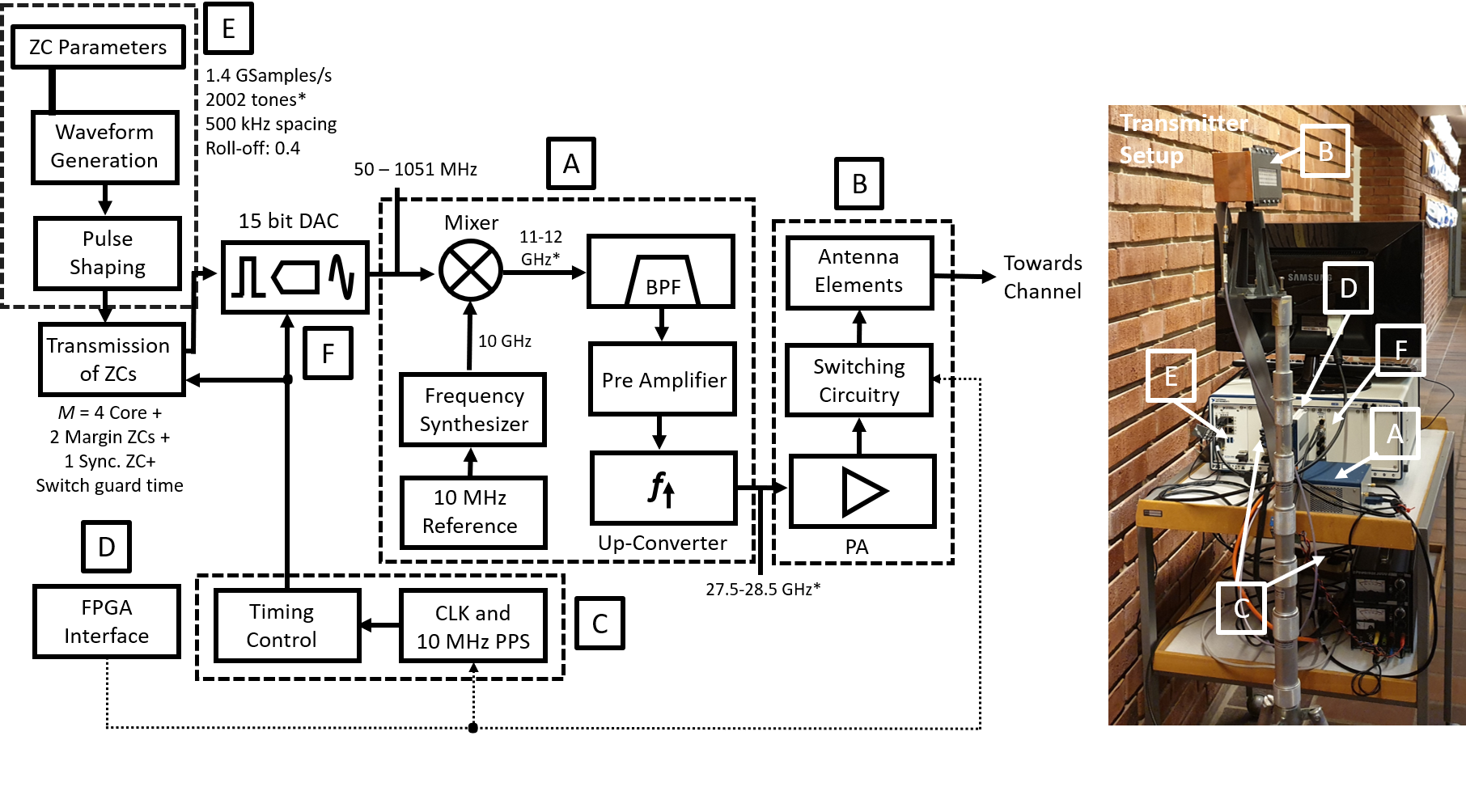}
    \vspace{-22pt}
    \caption{Left-hand side: Overall design of the transmitter hardware architecture for switched array operation between 27.5-29.5 GHz. The annotations ``A" to ``F" denote the various transmitter circuit components, mapping to their physical make up as shown on the right-hand side of the figure. These are further described in the text.  The terms ``PPS" and ``BPF" are read as ``pulse per-second" and ``bandpass filter", respectively.}
    \label{fig:TransmitterArchitecture}
\end{figure}

The transmitter is implemented and controlled in real-time with the National Instrument PXI-e 1085 chassis, supported by National Instrument's LabVIEW framework for software interfacing and control. The left-hand side of Fig.~\ref{fig:TransmitterArchitecture} shows the overall implementation of the transmitter hardware, where the physical circuit blocks are categorized and labelled from ``A" to ``F". Each of these correspond to the right-hand side of the figure where the realized transmitter setup is depicted. As it can be observed from the design of the transmitter, parts of its operations are controlled by the local host PC (see ``E" on the fig.), while others require integration with a FPGA (denoted with ``D"). The earlier explained frame structure (including the ZC sounding waveform with $M$ = 4) is generated at the host, and is shaped with root-raised-cosine filtering before interfacing to the FPGA for repetitive transmission. 
\begin{figure}[!t]
    \vspace{-5pt}
    \centering
    \hspace{3pt}
    \includegraphics[width=12.4cm]{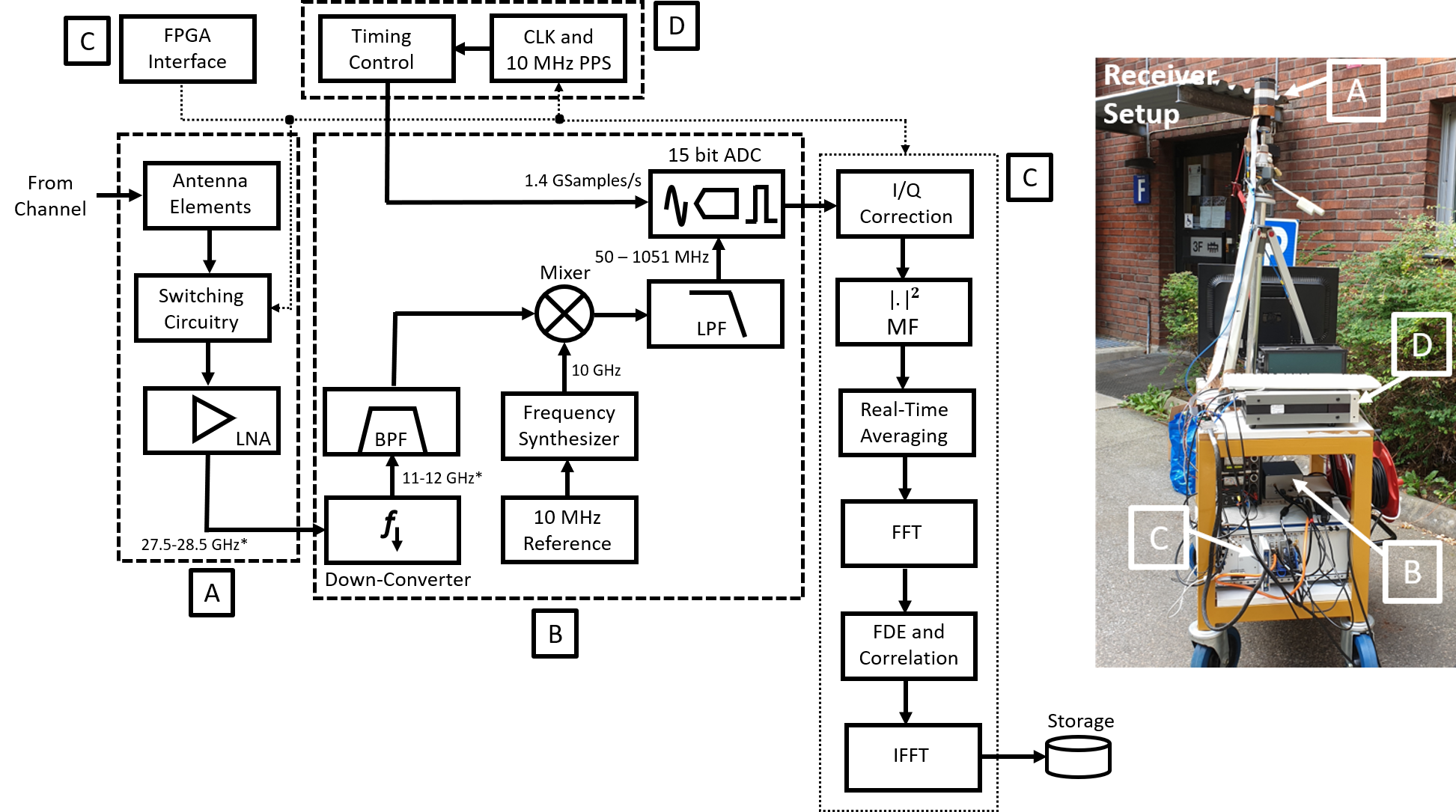}
    \vspace{-5pt}
    \caption{Left-hand side: Overall receiver hardware for switched array operation between 27.5-29.5 GHz. The annotations ``A" to ``D" represent the various receiver circuit components, mapping to their physical equivalents as shown on the right-hand side of the figure. The terms ``LPF", ``I/Q", ``MF", ``FFT", ``FDE" and ``IFFT" denote ``low-pass filter", ``in-phase/quadrature", ``matched filter", ``fast Fourier transform", ``frequency domain equalization", and ``inverse FFT", respectively.}
    \label{fig:ReceiverArchitecture}
    \vspace{-21pt}
\end{figure}
The transmission process triggers the timing control governed by the FPGA, which receives an external 10 MHz CLK reference (shown as ``C"). Similar to the designs in \cite{PAPAZIAN1,BAS1}, the CLK used for the sounding setup is based on a rubidium reference and is extremely stable providing CLK drifts on the order of 1 ns. A 15-bit digital-to-analog converter (DAC) output (shown as ``F") is then fed to the transmit RF head unit (marked with an ``A") for mixing the baseband signal to an IF of 11-12 GHz via the aid of a frequency synthesizer outputting 10 GHz, where it is filtered and pre-amplified before up-conversion to the desired band of 27.5-28.5 GHz. As marked by * on the left-hand side of the figure, where applicable, the parameters quoted in the design are for 1 GHz bandwidth (from 27.5 - 28.5 GHz). Despite this, the architecture is re-configurable to cover a 2 GHz bandwidth from 27.5 - 29.5 GHz when extending the parameters reported to cater for the bandwidth adjustment. This is also true when scaling down the bandwidth, to say 500 MHz. A separate PA is attached prior to the earlier described RF switching topology to give an additional gain of 28 dBm (especially since each switch yields a 1.5 dB insertion loss), while the CLK also triggers the switch timing (see ``B" on the figure). A control signal activates a random switch state according to the codebook of randomly designed switch entries known to both transmitter and receiver, forwarding the signal to the desired antenna port for transmission over-the-air. Taking into account the losses and power gains at each stage, an effective isotropic radiated power (EIRP) of approximately 43 dBm is achieved.\footnote{Due to the complexity of the system and its many involved blocks, we do not attempt to further breakdown the EIRP of 43 dBm into the various smaller contributions from individual components as this deviates away from the main focus of the paper. Also, our definition of the EIRP \emph{includes} the gain harvested from the coherent post-processing step (presented later in the text).} Later we provide specific part details of the used circuit blocks. 

The corresponding architecture at the receiver is shown on the left-hand side of  Fig.~\ref{fig:ReceiverArchitecture}. In accordance with the same pre-defined codebook as for the transmitter, the receive antenna elements and its associated switch circuitry is controlled via the CLK reference signal which triggers the capturing of the received waveform. This is followed by a LNA, and a down-converter to IF before being further mixed and low pass filtered (LPF) to baseband. A 15-bit analog-to-digital converter (ADC) is then employed to digitize the receive analog waveform. The receiver FPGA is then used to perform in-phase/quadrature (I/Q) sample correction followed by a power maximization stage introduced by the matched filter (MF). The FPGA performs real-time averaging of the many acquired replicas of ZC sequences (which were transmitted) and transforms the averaged waveform to the frequency domain via fast Fourier transforms (FFTs) for equalization and further processing. The resulting average is \emph{correlated} with the transmitted ZC sequence and converted back to time domain via inverse FFTs (IFFT) to extract the influence of the propagation channel. The channel impulse response is then extracted and is sent for storage (with a streaming rate of 1 Gb/s) and further post-processing to extract the directional propagation parameters. \emph{Due to the large number of elements being switched at both the transmitter and receiver (32768 combinations in a snapshot), we stress that the above process is very complex and requires optimization of both hardware as well as real-time software.}\footnote{The post-processing also involves calibration and addition of metadata, e.g., absolute timing and antenna pair identification, which are not mentioned for the sake of maintaining clarity.} The terms ``A" to ``D" on the left-hand side of the fig. categorize the system components and are depicted physically on the right-hand side where the complete receiver setup is shown. The receiver hardware is also interfaced with a National Instrument PXI-e 1085 chassis, controlled by the custom designed host and FPGA LabVIEW software frameworks. For both the transmitter and receiver, the different utilized circuit parts are provided in Tab.~\ref{Tab:CircuitElements}. 
\begin{table}[!t]
\vspace{-5pt}
\begin{centering}
\scalebox{0.6}{
\begin{tabular}{ccc}
\toprule 
\textbf{Part Number} & \textbf{System Unit} & \textbf{Manufacturer and Model Number}
\tabularnewline
\midrule
\midrule 
1 & PA & Sage Millimeter Inc., SBP-2633332228-KFKF-S1 \tabularnewline
\midrule 
2 & LNA & Pasternack Inc., PE15A3300 \tabularnewline
\midrule 
3 & Transmitter RF Head & National Instruments 3642\tabularnewline
\midrule
4 & Receiver RF Head & National Instruments 3652\tabularnewline
\midrule
5 & LO and IF Blocks & National Instruments PXI-e 3620\tabularnewline
\midrule
6 & DAC & National Instruments PXI-e 3610\tabularnewline
\midrule
7 & ADC & National Instruments PXI-e 3630\tabularnewline
\midrule
8 & FPGAs & National Instruments PXI-e 7976 R \tabularnewline
\midrule
9 & Switch Control & National Instruments 6581 B\tabularnewline
\midrule
10 & Timing and Synchronization Modules & National Instruments PXI-e 6674 T\tabularnewline
\midrule
11 & Host PC Framework & National Instruments PXI-e 8880\tabularnewline
\midrule
12 & Chassis Control & National Instruments PXI-e 1085 \tabularnewline
\midrule
13 & Fast Array Hard Drive for Storage & National Instrument PXI-e HDD 8261\tabularnewline
\midrule
14 & CLK Source & Stanford Research Systems Rubidium Frequency Standard FS725 \tabularnewline
\bottomrule
\end{tabular}}
\vspace{11pt}
\caption{Transmitter and Receiver system components of the designed channel sounder. Unless otherwise stated, the listed components apply to both link ends.}
\label{Tab:CircuitElements}
\par\end{centering}
\vspace{-41pt}
\end{table}
\subsubsection{Key Real-Time Software Implementation Aspects}
\label{KeyRealTimeSoftwareImplementationAspects}
From a real-time software viewpoint, both the transmit and receive sides have a similar overall code framework. The code is divided in \emph{three} parts: (1) Host code; (2) FPGA signal processing code, and (3) FPGA switch control code. The host code is the main controller, and acts as a user interface at both sides. It mainly controls the operational states and executes: (1) The ZC waveform, (2) The number of ZC repetitions, and (3) The number of switch states, i.e., the codebook. At the receiver, it additionally receives the raw channel impulse responses and adds the required metadata (absolute times, antenna pair indices, etc.) before saving. Unlike the host implementation, the signal processing FPGA code is clocked by the rubidium reference. The ZC signals are transmitted with the pre-defined number of repetitions, and the switch guard times are added together with a switch triggering signal which is generated between each frame. The receive side, similarly, processes each frame, generates a switch triggering signal across the received frames and sends the resulting channel impulse response with a timing identifier added to the host. The aforementioned processing involves synchronization, averaging and correlation of the received ZC waveform to a known ZC signal. The switch FPGA code is the least complex of the three, and remains the same at both link ends. Its main role is to change the switch states based on the pre-defined codebook (defining the pseudo-random switching combinations) each time it receives a trigger signal from the real-time signal processing FPGA code. With the major aspects of the sounder design concluded, we now present the final specifications of the sounder and assess the sounder's dynamic range in order to understand its maximum measurable range.  
\vspace{-14pt}
\subsection{Final Sounder Specifications and Dynamic Range}
\label{FinalSounderSpecificationsandDynamicRange}
\vspace{-3pt}
\begin{table}[!t]
\begin{centering}
\scalebox{0.58}{
\begin{tabular}{cc}
\toprule 
\textbf{Parameter}&\textbf{Value} 
\tabularnewline
\midrule
\midrule 
Sounder type & Switched array \tabularnewline
\midrule
Center frequency  & 28 GHz (re-configurable*) \tabularnewline
\midrule 
Instantaneous bandwidth & Up to 2 GHz (re-configurable*) \tabularnewline 
\midrule
Antenna array sizes/configurations & 128 by 256/UPA and cylindrical (transmit and receive) \tabularnewline
\midrule
Azimuth/Elevation 3 dB beamwidths & 85$^{\circ}$/50$^{\circ}$\tabularnewline
\midrule
Overall field-of-view at transmitter/receiver & 180$^{\circ}$/360$^{\circ}$\tabularnewline
\midrule
Polarization & Dual (horizontal and vertical) (re-configurable*) \tabularnewline
\midrule
Transmit EIRP & 43 dBm \tabularnewline
\midrule
Receive array gain & 30.08 dB \tabularnewline
\midrule
Receiver noise figure & 5 dB \tabularnewline 
\midrule
Maximum PA/LNA gains & 32 dBm/43 dBm\tabularnewline 
\midrule
ADC/DAC resolution and sampling rate & 15 bits at 1.4 GSamples/s\tabularnewline
\midrule
Switching rate & 18.3 $\mu$s (re-configurable*)\tabularnewline 
\midrule
MIMO snapshot rate & Around 600 ms (re-configurable*)\tabularnewline
\midrule
Switching sequence & Pseudo-random via a codebook (re-configurable*)\tabularnewline
\midrule
Transmitter-receiver CLK synchronization & Rubidium reference\tabularnewline
\midrule
Transmitted waveform duration & 2.6 $\mu$s \tabularnewline
\midrule
Number of tones & 2002 (re-configurable*)\tabularnewline
\midrule
Tone spacing in frequency & 500 kHz (re-configurable*)\tabularnewline
\midrule 
PAPR & 0.349 dB \tabularnewline 
\bottomrule 
\end{tabular}}
\vspace{10pt}
\caption{Overall specifications of the designed channel sounder. The parameters which can be re-configured are listed with ``(re-configurable*)".}
\label{Tab:FinalSpecifications}
\par\end{centering}
\vspace{-40pt}
\end{table}
The overall specifications of the sounder are given in Tab.~\ref{Tab:FinalSpecifications}. As shown by the tabulated entries, our design offers an unprecedented degree of re-configurability which can be exercised in-between measurements without any changes to the hardware. In particular, the center carrier frequency, bandwidth, total number of measured channels, number of active polarization modes, the total number of ZC sequences, and the switching sequence configuration can be re-configured depending on the measurement campaign. This enables, for instance, to get faster switching rates in either azimuth/elevation for one polarization mode for high mobility situations, where we know that MPCs are highly dynamic or mainly bounded within a certain angular range incident in the horizontal/vertical plane without significant influence of polarization diversity. In contrast, for more static channels, a longer measurement can be made for getting a detailed description of the channel conditions. This provides a considerable advantage relative to existing setups summarized in Sec.~\ref{Introduction} of the paper, which cannot be re-configured instantaneously without extensive mechanical modifications to the sounder structure and re-cabling.

\begin{figure}[!t]
\vspace{-15pt}
    \centering
    \includegraphics[width=9cm]{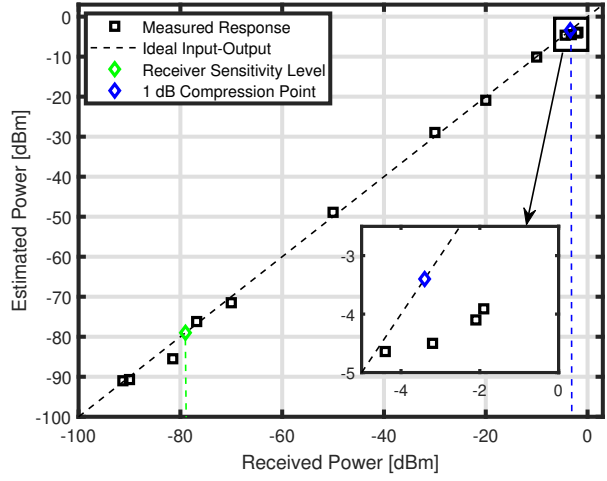}
    \vspace{-5pt}
    \caption{Measured dynamic range characteristics of the designed channel sounder.}
    \label{fig:DynamicRangeMeasured}
    \vspace{-20pt}
\end{figure}

Cascading the individual noise figures of the circuit components shown in Fig.~\ref{fig:ReceiverArchitecture}, the estimated noise figure for the receiver is around 5 dB. From this, with 1 GHz measurement bandwidth (naturally extendable to 2 GHz), the receiver sensitivity level can be computed by
\vspace{-6pt}
\begin{equation}
    \label{receiversensitivity}
    \textrm{Receiver Sensitivity}=-174\hspace{3pt}\textrm{dBm/Hz} + 5 \hspace{3pt}\textrm{dB} + 10\log_{10}\left(1\times{}10^{9}\hspace{3pt}\textrm{Hz}\right)=-79\hspace{3pt}\textrm{dBm}.
    \vspace{-8pt}
\end{equation}
Figure~\ref{fig:DynamicRangeMeasured} depicts the \emph{measured} dynamic range of the sounder. The received power on the horizontal axis is the calculated power given the transmit EIRP harvested in the coherent processing step. We note that the different power levels are achieved by varying the raw transmit power. The estimated power from the recorded waveform averaged over the whole band is depicted on the vertical axis. Together with the estimated receiver sensitivity, the noise power distorts the estimated power if the input power is below -79 dBm. In addition, the receiver starts to saturate around -4 dBm (see the zoomed part of the figure). Hence, the channel sounders dynamic range is 75 dB. We recognize that the receiver sensitivity formulation in \eqref{receiversensitivity} does not include the receive array gain, $G_{\textrm{receive}}$, also with coherent post-processing. As such, the equivalent isotropic sensitivity level can be given by $\textrm{Receiver Sensitivity}-G_{\textrm{receive}}=-79\hspace{3pt}\textrm{dBm}-(30.08\hspace{3pt}\textrm{dB})=-109.08\hspace{3pt}\textrm{dBm}$ \cite{ZHANGA}. Together with the 43 dBm of EIRP (harvested with post-processing), the maximum measurable pathloss with the designed sounder will be around 152 dB. Despite a large number of channels being measured within short times with both high spatial and temporal resolutions, this compares well with existing sounders reported in \cite{BAS1,SUN1,GENTILE1,SALOUS1}. The state-of-the-art rotating horn antenna sounder in \cite{MACARTNEY1} can measure pathloss up to 185 dB, at the expense of much larger measurement durations since its design is based on a sliding correlator, and is thus not suitable for dynamic channel characterization. If the same acquisition time of 32.752 ms is considered for sounding one transmit-receive antenna combination while performing 20 averages of the received waveform, with our design, we can repeat the same measurement 233$\times$ which would increase the maximum measurable pathloss well beyond 185 dB due to the gain brought by the equivalently high number of complex averages. Due to space constraints, we do not present explicit pathloss vs. distance evaluations, since these are, by now, well characterized in the literature \cite{BAS1,SUN1,GENTILE1,SALOUS1,MACARTNEY1}. Instead, in Sec.~\ref{SampleMeasurementResults}, our focus is on extracting directional MPC parameters for both dynamic and static scenarios. We now summarize the post-processing routine for extracting directional MPC parameters. 
\vspace{-12pt}
\section{Post-Processing Aspects}
\label{PostProcessingAspects}
\vspace{-3pt}
\subsection{Channel Impulse Response Determination}
\label{ChannelImpulseResponseDetermination}
The transfer function of the measured propagation channel can be represented as a fourth order tensor denoted by $\mathbf{H}\hspace{1pt}(s,t,f,r)$ from which the spatio-temporal information of the MIMO channel can be derived. Here $s$ denotes the measured snapshot index, $f$ is the measured frequency index with $f=1,2,\dots,M_f$ denoting the number of frequency points, $t=1,2,\dots,M_T$ is the transmit antenna index and $r=1,2,\dots,M_R$ is the receive antenna index, respectively. Naturally, each snapshot contributes to the improvement in measurement SNR by providing robustness against measurement noise. The channel impulse response is then denoted by $\mathbf{h}\hspace{1pt}(s,\tau,t,r)$, where $\tau$ denotes the delay and is obtained via an inverse Fourier transform with a Hann window to supress sideband aliasing. To this end, the non-directional local power-delay profile (PDP) is computed as $\textrm{PDP}(s,\tau,t,r)=|\mathbf{h}\hspace{1pt}(s,\tau,t,r)|^{2}$. 

\vspace{-15pt}
\subsection{Directional MPC Parameter Extraction}
\label{DirectionalMPCParameterExtraction}
\vspace{-2pt}
Utilizing the dense spatial and temporal resolutions on offer with the designed sounder, our ultimate interest is to obtain propagation channel characteristics that are \emph{independent} of the antenna architectures, such that the true directional properties of only the MPCs can be evaluated. Naturally, to do this, we would need to obtain a double-directional characterization of the channel that extracts the spatio-temporal polarimetric parameters of the MPCs from the transfer function via the use of a high resolution estimation algorithm. Several approaches exist in the literature to do this, such as space alternating generalized expectation maximization (SAGE) \cite{FLUREY1}, CLEAN \cite{SCOTT1} and RIMAX \cite{RICHTER1,TATARIA2,SANG1}. All of these approaches attempt to extract the directional MPC parameters in the maximum likelihood sense. These approaches are subjected to different model assumptions and underlying conceptual restrictions including applicability to certain antenna array architectures, calculation time in terms of convergence speed, and statistical efficiency. It is now well known that the SAGE and CLEAN algorithms can be applied to a large range of antenna array architectures, with the drawback of slow convergence rates if two closely spaced propagation paths exist in the channel. Furthermore, the interaction of the specular/quasi-deterministic MPCs and more diffuse MPCs seems to be missing from the SAGE and CLEAN approaches. These issues are overcome in RIMAX - an iterative maximum-likelihood estimator, which results in a more detailed description of the directional parameters \cite{RICHTER1}. As such, we employ the use of RIMAX and provide a methodology for obtaining MPC parameters. 

With RIMAX, the propagation channel  
\vspace{-7pt}
\begin{equation}
    \label{MPCExtraction1}
    \mathbf{h}=\mathbf{S}(\theta_{\textrm{specular}})+\mathbf{D}(\theta_{\textrm{dense}})+\mathbf{n} \in \mathbb{C}^{M_{T}M_{R}M_{f}\times{}1},
    \vspace{-5pt}
\end{equation} 
is considered as a superposition of the specular MPCs, denoted by $\mathbf{S}(\theta_{\textrm{specular}})$, dense MPCs, denoted by $\mathbf{D}(\theta_{\textrm{dense}})$ and the measurement noise, denoted by  $\mathbf{n}$. 
The specular MPCs are characterized via plane waves through their delays ($\tau$), azimuth and elevation AOAs ($\phi_{\hspace{1pt}\textrm{R}}$ and $\varphi_{\hspace{1pt}\textrm{R}}$), azimuth and elevation AODs ($\phi_{\hspace{1pt}\textrm{T}}$ and $\varphi_{\hspace{1pt}\textrm{T}}$), and complex polarimetric path gains ($\gamma$), such that  $\theta_{\textrm{specular}}=[\tau,\phi_{\textrm{R}},\varphi_{\textrm{R}},\phi_{\textrm{T}},\varphi_{\textrm{T}},\gamma]$. Analytically, 
\begin{equation}
\label{MPCExtraction2}
\mathbf{S}\hspace{-1pt}\left(\theta_{\textrm{specular}}\right)=\sum\limits_{\ell=1}^{L}\mathbf{B}_{\textrm{R}}^{T}\left(\phi_{\textrm{R},\ell},\varphi_{\textrm{R},\ell}\right)
\begin{pmatrix}
\gamma_{\textrm{HH},\ell} & \gamma_{\textrm{HV},\ell}\\
\gamma_{\textrm{VH},\ell} & \gamma_{\textrm{VV},\ell}
\end{pmatrix}
\mathbf{B}_{\textrm{T}}\left(\phi_{\textrm{T},\ell},\varphi_{\textrm{T},\ell}\right)e^{-j2\pi{}f\tau_{\ell}}. 
\vspace{-1pt}
\end{equation}
From \eqref{MPCExtraction2} it is clear that there are $L$ specular MPCs which are to be estimated and $(\cdot)^T$ denotes the matrix transpose operator. Note that $\mathbf{B}_{\textrm{R}}$ and $\mathbf{B}_{\textrm{T}}$ denote the \emph{non-linear projection matrices} mapping the azimuth and elevation AOAs and AODs (for each specular MPC) to the antenna array responses obtained from the array calibration measurements discussed in Sec.~\ref{ArrayCalibration}. This mapping is conducted via design of the effective aperture distribution function (EADF) \cite{LANDMANN1,RICHTER1} of both transmit and receive arrays. The EADF is a known procedure in measurement-based parameter estimation, which is computed via a two-dimensional fast Fourier transform and subsequent data compression of the precisely measured complex polarimetric gain and phase patterns of both arrays. Furthermore, the parameters  $\gamma_{\textrm{HH},\hspace{1pt}\ell}$, $\gamma_{\textrm{HV},\hspace{1pt}\ell}$, $\gamma_{\textrm{VH},\hspace{1pt}\ell}$ and $\gamma_{\textrm{VV},\hspace{1pt}\ell}$ denote the horizontal-to-horizontal, horizontal-to-vertical, vertical-to-horizontal and vertical-to-vertical polarization gains of MPC $\ell$. We note that $\tau_\ell$ denotes the time delay associated with the $\ell$-th specular MPC. The dense MPCs describes the stochastic part of the channel and is assumed to comprise of a large number of individually weak MPCs that can not be individually estimated as plane waves due to the underlying physical process differences in diffuse scattering, wavefront curvature effects, etc. Owing to the central limit theorem, $\mathbf{D}({\theta_{\textrm{dense}}})$ is modelled as a zero-mean complex circularly symmetric Gaussian vector with covariance defined by $\mathbf{R}_{\hspace{1pt}\textrm{dense}}\in\mathbb{C}^{M_{T}M_{R}M_{f}\times{}M_{T}M_{R}M_{f}}$. Thus, $\mathbf{D}(\theta_{\textrm{dense}})\sim\mathcal{CN}(0,\mathbf{R}_{\textrm{dense}})$. The measurement noise (including thermal and ambient noise) is also assumed to have a complex Gaussian structure, such that $\mathbf{n}\sim\mathcal{CN}(0,\sigma_{\textrm{n}}^2\hspace{1pt}\mathbf{I})$ with variance $\sigma_{\textrm{n}}^2$. Following the approach in \cite{RICHTER1}, we estimate the dense MPCs in conjunction with the measurement noise resulting in an aggregate covariance matrix $\mathbf{R}_{\textrm{aggregate}}=\mathbf{R}_{\textrm{dense}}+\sigma_{\textrm{n}}^2\hspace{1pt}\mathbf{I}$. This enables us to write  \eqref{MPCExtraction1} as 
\vspace{-14pt}
\begin{equation}
    \label{MPCExtraction3}
    \mathbf{h}=\mathbf{S}\left(\theta_{\textrm{dense}}\right)+
    \mathbf{n}_{\textrm{aggregate}}\in\mathbb{C}^{M_{T}M_{R}M_{f}\times{}1}, 
    \vspace{-8pt}
\end{equation}
where $\mathbf{n}_{\textrm{aggregate}}\sim\mathcal{CN}(0,\mathbf{R}_{\textrm{aggregate}})$. Although more sophisticated methods have been used \cite{KASKE1} to model the dense MPCs, we adhere to the model employed in \cite{SANG1,KASKE2,SALMI1}. As such, our analysis is performed over an \emph{aggregate} of the different polarization components. The co-variance matrix can be further decomposed into the Kronecker product of three other matrices
\vspace{-7pt}
\begin{equation}
    \label{MPCExtraction4}
    \mathbf{R}_{\textrm{dense}}=\mathbf{R}_{\textrm{R}}\hspace{1pt}\otimes\mathbf{R}_{\textrm{T}}\hspace{1pt}\otimes\mathbf{R}_{\textrm{F}},
    \vspace{-9pt}
\end{equation}
where $\mathbf{R}_{\textrm{F}}$ is the covariance matrix in the frequency domain, and $\mathbf{R}_{\textrm{R}}$ and $\mathbf{R}_{\textrm{T}}$ are the covariance matrices of antenna elements at the transmitter and receiver, respectively. The frequency domain covariance matrix is modelled by $\mathbf{R}_{\textrm{F}}=\textrm{Toep}(\lambda\hspace{1pt}(\theta_{\textrm{F}}),\lambda\hspace{1pt}(\theta_{\textrm{F}})^{H})$, where $\textrm{Toep}(\cdot)$ and $(\cdot)^H$ denote a Toeplitz matrix and matrix Hermitian transpose, respectively. Note that $\lambda(\theta_{\textrm{F}})$ is a \emph{sampled} version of the power spectral density given by 
\vspace{-4pt}
\begin{equation}
    \label{MPCExtraction5}
    \lambda\left(\theta_{\textrm{F}}\right)=\frac{\bar{\gamma}_{1}}{M_{\textrm{F}}}\left(\frac{1}{\beta_{\textrm{d}}},\frac{e^{-j2\pi\tau_{\textrm{d}}}}{\beta_{\textrm{d}}+j2\pi\frac{1}{M_{f}}},\dots{},\frac{e^{-j2\pi\left(M_{f}-1\right)\tau_{\textrm{d}}}}{\beta_{\textrm{d}}+j2\pi\frac{M_{f}-1}{M_{f}}}\right)^{\hspace{-3pt}T}\hspace{-3pt}. 
    \vspace{-3pt}
\end{equation}
The parameters of the frequency domain covariance matrix model are $\theta_{\textrm{F}}=[\tau_{\textrm{d}},\beta_{\textrm{d}},\bar{\gamma}_{1}]^{T}$, where $\tau_{\textrm{d}}$ and $\bar{\gamma}_{1}$ are the time delay of arrival and power of the first component in the time domain version of 
$\mathbf{R}_{\textrm{F}}=\textrm{Toep}\hspace{1pt}(\lambda\hspace{1pt}(\theta_{\textrm{F}}),\lambda\hspace{1pt}(\theta_{\textrm{F}})^{H})$. On the other hand, $\mathbf{R}_{\textrm{T}}$ and $\mathbf{R}_{\textrm{R}}$ (denoted by $\mathbf{R}_{\textrm{R}/\textrm{T}}$ in this explicit case) are modelled as $\mathbf{R}_{\textrm{R/T}}=\mathbf{B}_{\textrm{R/T}}
    \hspace{2pt}\mathbf{\Delta}\hspace{-2pt}\left(\theta_{\textrm{R/T}}\right)\mathbf{B}_{\textrm{R/T}}^{H}$, where $\mathbf{B}_{\textrm{R/T}}$ captures the array responses at the receiver or transmitter, while $\mathbf{\Delta}(\theta_{\textrm{R/T}})$ is a diagonal matrix with entries determined by the \emph{angular} (jointly in azimuth and elevation) probability density function of the dense MPCs. As motivated and proven in \cite{KASKE2}, this is best represented with a von Mises distribution. Following the process outlined in \cite{KASKE2}, the parameters of $\mathbf{\Delta}(\theta_{\textrm{R/T}})$ are given by $\theta_{\textrm{R}}=[\mu_{\phi,\textrm{R}},\mu_{\varphi,\textrm{R}},\kappa_{\phi,{\textrm{R}}},\kappa_{\varphi,\textrm{R}},\tilde{\gamma}_{\phi,\textrm{R}},\tilde{\gamma}_{\varphi,\textrm{R}}]$ and  $\theta_{\textrm{T}}=[\mu_{\phi,\textrm{T}},\mu_{\varphi,\textrm{T}},\kappa_{\phi,{\textrm{T}}},\kappa_{\varphi,\textrm{T}},\tilde{\gamma}_{\phi,\textrm{T}},\tilde{\gamma}_{\varphi,\textrm{T}}]$. Here $\mu$, $\kappa$ and $\tilde{\gamma}$ are the mean AOAs/AODs, concentration parameters and amplitude coefficients in the azimuth or elevation plane. For further discussions, we refer the interested reader to \cite{RICHTER1,SANG1,KASKE2,SALMI1} and references therein.

\vspace{-12pt}
\section{Sample Measurement Results}
\label{SampleMeasurementResults}
\vspace{-5pt}
\subsection{Measurement Scenarios}
\vspace{-2pt}
\label{MeasurementScenarios} 
\begin{figure}[!t]
   \vspace{-13pt}
    \centering
    \includegraphics[width=10cm]{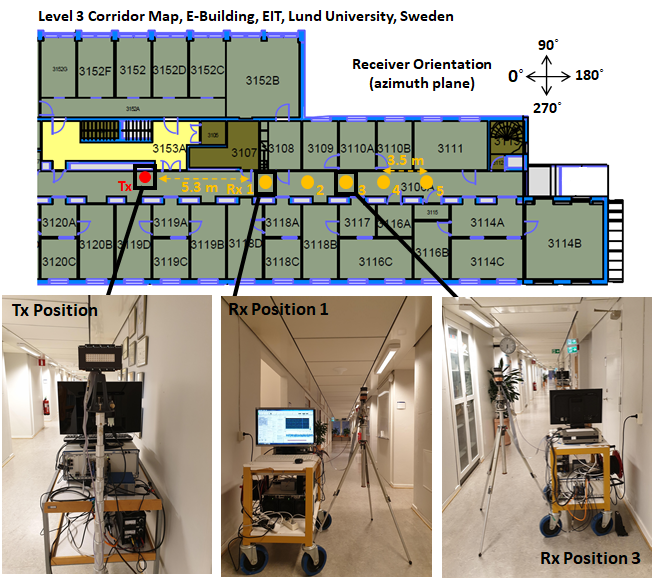}
    \vspace{-5pt}
    \caption{Geometry of the static propagation measurements in a typical indoor office environment with the locations of the transmitter (``Tx") and receiver (``Rx") marked out along with the relative distances. The surroundings of the transmitter, and receive positions 1 and 3 are shown.}
    \label{fig:StaticMeasurementScenario}
    \vspace{-3pt}
\end{figure}
\begin{figure}[!t]
   \centering
   \vspace{-5pt}
    \includegraphics[width=10cm]{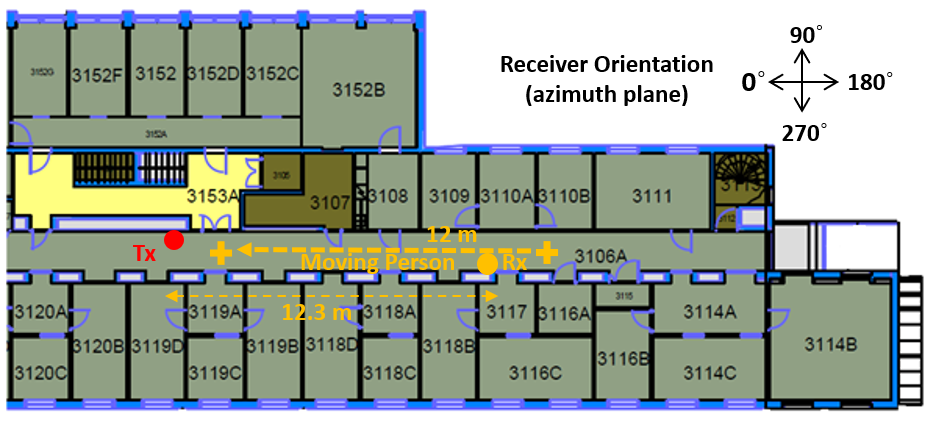}
    \vspace{-8pt}
    \caption{A description of the dynamic measurement scenario with the transmitter, receiver locations as well as the trajectory of the moving person in between the transmitter and receiver.} 
    \label{fig:DynamicMeasurementScenario}
    \vspace{-25pt}
\end{figure}
To validate the performance of the designed sounder, we present results from an indoor measurement campaign carried out at Level 3 of the E-Building, Department of Electrical and Information Technology (EIT), Lund University, Sweden. The environment can be described as a typical office corridor with a width of 2.4 m and a height of 2.6 m. The campaign covered both static and dynamic scenarios, where the precise geometry for the static scenario is as depicted in Fig.~\ref{fig:StaticMeasurementScenario}. The transmitter position is marked with a red dot (labelled as ``Tx"), from which the channel is measured at five receive locations marked with yellow circles (and labelled as ``Rx 1,2,3,4,5") down the corridor. The transmitter and receiver heights were 1.57 m, respectively and the surroundings of the transmitter and receive positions 1 and 3 are depicted in Fig.~\ref{fig:StaticMeasurementScenario}. Due to it being a static measurements, movement in the channel was not present, and naturally, the both link ends were fixed at the positions depicted in Fig.~\ref{fig:StaticMeasurementScenario}. At each receive position, 5 snapshots were averaged and later analyzed. In addition to the static, a dynamic scenario was also measured as shown in Fig.~\ref{fig:DynamicMeasurementScenario}. Here, the receiver was fixed at 12.3 m from the transmitter location, while a person was moving at approximately 1 m/s on the trajectory shown in the yellow dashed line in between the transmitter and receiver. Since around 2 snapshots per-second were measured, a total of 24 snapshots were acquired. For both scenarios, the measurements were conducted using all antennas at both link ends (i.e., measuring 32768 channel combinations) across 27.5-28.5 GHz with a center frequency of 28.0 GHz. All directional MPC extraction was conducted using the process outlined in Sec.~\ref{DirectionalMPCParameterExtraction}. The approach articulated in \cite{KASKE2} was employed to get around ``false" MPCs.  
\vspace{-15pt}
\subsection{Measured Results and Evaluations} 
\label{MeasuredResultsandEvaluations}
\begin{figure}[!t]
   \centering
   \vspace{-10pt}
    \includegraphics[width=15cm]{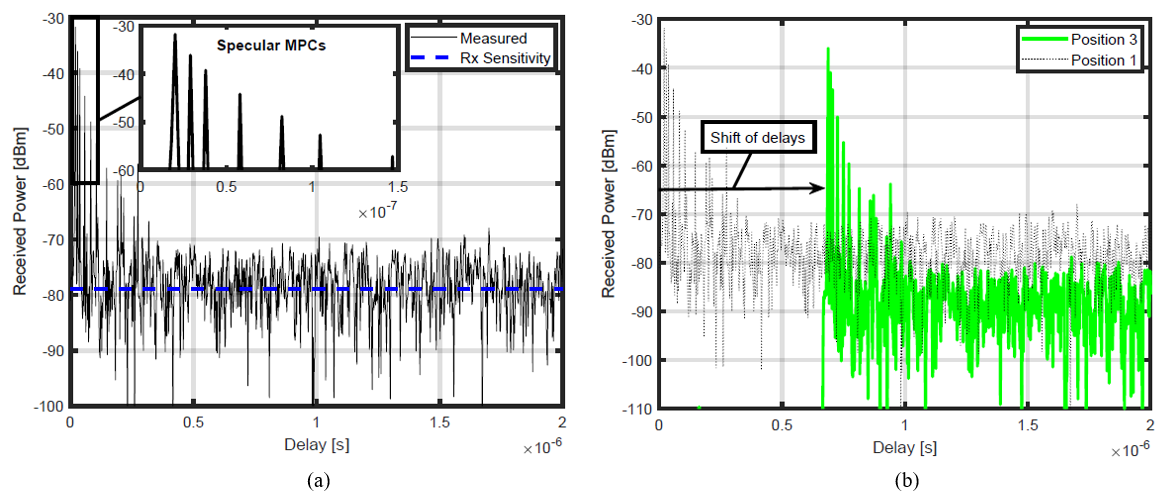}
    \vspace{-16pt}
    \caption{Directional PDPs at receiver positions (a) 1 and (b) 3 for a static measurement scenario described in Fig.~\ref{fig:StaticMeasurementScenario}.}
    \label{fig:StaticDDPs}
    \vspace{-22pt}
\end{figure}

Figure~\ref{fig:StaticDDPs} (a) and (b) depict the directional PDPs at receiver positions 1 and 3, respectively. At position 1, one can notice an early cluster of specular MPCs at the beginning of the PDP (see the zoomed part of the figure). These mostly arise from line-of-sight (LOS) and specular reflections around LOS, including from the corridor walls either side of the receiver. With increasing MPC delay, power drops off in the classical exponentially decaying manner. Reflections from the ceiling may also visible due to the considerably wide elevation look angles of both the transmit and receive array. Some dominant diffuse MPCs are also visible just after the LOS cluster in addition to many weaker diffuse components. The receiver noise sensitivity level is shown in blue dashed line. Sub-figure (b) demonstrates the directional PDP at receiver position 3, where one can clearly see a shift in the MPC delays along with a degradation in the directional received power as a function of the increased distance from the transmitter to the receiver position. 
\begin{figure}[!t]
\vspace{-15pt}
    \centering
    \hspace{-12pt}
    \includegraphics[width=9.2cm]{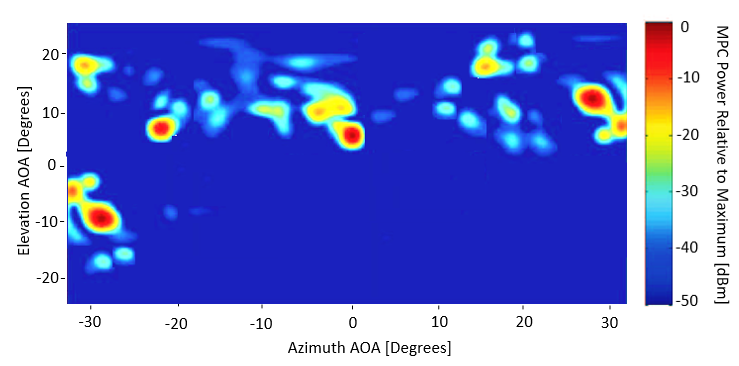}
    \vspace{-10pt}
    \caption{Power angular spectrum at receiver position 1 considering the static measurement scenario.}
    \label{fig:PASStaticPos1}
    \vspace{-3pt}
\end{figure}
\begin{figure}[!t]
    \centering
    \vspace{-14pt}
    \includegraphics[width=9.5cm]{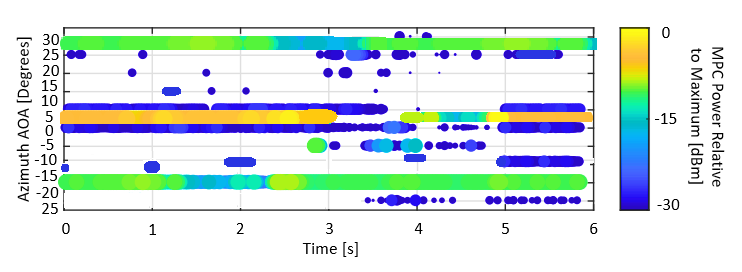}
    \vspace{-15pt}
    \caption{Dynamic (time varying) azimuth AOA as a function of power and time for scenario presented in Fig.~\ref{fig:DynamicMeasurementScenario}.}
    \label{fig:DynamicAzimuthalAOAs}
    \vspace{-23pt}
\end{figure}
Figure~\ref{fig:PASStaticPos1} demonstrates the power-angular spectrum in both the azimuth and elevation domains when the receiver is located at position 1. Here in accordance with the directional PDP shown in Fig.~\ref{fig:StaticDDPs}(a), individual specular and diffuse MPCs can be resolved from an azimuth angular range of 60$^\circ$, while a relatively larger spread of around 40$^{\circ}$ in elevation is also visible. The high resolution of the PAS is readily visible, which enables us to study the propagation behavior in rather precise detail. In contrast to Figs.~\ref{fig:StaticDDPs} and \ref{fig:PASStaticPos1}, Fig.~\ref{fig:DynamicAzimuthalAOAs} depicts the dynamic directional azimuth AOA as a function of time and power, considering the scenario presented in Fig.~\ref{fig:DynamicMeasurementScenario}. Analyzing a total time horizon of 6 s (12 snapshots), the azimuth AOAs are tracked over both time and space with high resolution, where the appearance/disappearance effects of the AOAs is readily visible. Between time 3 and 4 s, blockage of the LOS component can be observed due to the moving person passing in front of the receiver setup. Specular MPCs (in the form of dominant AOAs) from the corridor walls are also seen at azimuth AOA of -15$^{\circ}$, -5$^{\circ}$ and 30$^{\circ}$, respectively. Weaker azimuth AOAs (30 dB below the MPC with maximum power) are also visible scattered throughout the figure. Such results confirm the benefits of having high spatial and temporal resolutions on top of high resolution post-processing for fast, yet precise channel characterization. Naturally, many further details from the measured results can be presented. However, we refrain from doing this, since the central focus of the paper is on the design, implementation and measurement-based verification of the proposed sounder.

\vspace{-17pt}
\section{Conclusions}
\label{Conclusions}
\vspace{-5pt}
We presented a novel 27.5-29.5 GHz propagation channel sounder that can perform directional measurements in dynamic (and static) environments. Our sounder is based on the switched array principle, and is capable of characterizing 128$\times$256 dual-polarized channels within a short measurement time of around 600 ms with an underlaying antenna switching rate of 18 $\mu$s for 4 complex waveform averages. Unlike previously presented mmWave sounders, our design offers the highest degree of re-configurability and compliments the high delay resolution with spatial characteristics provided by the uniquely designed transmit and receive arrays. The shorter measurement times and stable RF design has allowed us to leverage post-processing gain to increase the received SNR and dynamic range. Using RIMAX, we show sample results which can characterize the directional MPC properties across both space and time facilitating accurate static and dynamic characterization.

\vspace{-20pt}
\bibliographystyle{IEEEtran}

\end{document}